\documentclass[useAMS,usenatbib,usegraphicx]{mn2e}

\usepackage{graphicx}  
\usepackage[applemac]{inputenc}
\usepackage{t1enc}
\usepackage{hyperref}    

\title[Exploring the spreading layer of GX 9+9 \ldots]{Exploring the spreading layer of GX 9+9 using \emph{RXTE} and \emph{INTEGRAL}}
\author[P. Savolainen et al.]{P. Savolainen$^{1}$\thanks{E-mail:
petri.savolainen@tkk.fi (PS), diana@kurp.hut.fi (DCH), osmi.vilhu@helsinki.fi (OV)}, D. C. Hannikainen$^{1}$, O. Vilhu$^{2}$, A. Paizis$^{3}$, J. Nevalainen$^{2}$ and
\newauthor P. Hakala$^{4}$\\
$^{1}$Metsähovi Radio Observatory, TKK, Metsähovintie 114, FIN-02540 Kylmälä, Finland\\
$^{2}$Observatory, PO Box 14, FIN-00014 University of Helsinki, Finland\\
$^{3}$INAF - Instituto di Astrofisica Spaziale e Fisica Cosmica, Sezione di Milano, Milano, Italy\\
$^{4}$Tuorla Observatory, University of Turku, FIN-21500 Piikkiö, Finland}
\begin{document}

\date{Accepted 2008 November 03. Received 2008 October 31; in original form 2008 July 17}

\pagerange{\pageref{firstpage}--\pageref{lastpage}} \pubyear{2008}

\maketitle

\label{firstpage}

\begin{abstract}

We have fitted $\sim200$ \emph{RXTE} and \emph{INTEGRAL} spectra of the neutron star LMXB GX 9+9 from 2002--2007 with a model consisting of a disc blackbody and another blackbody representing the spreading layer (SL), i.e. an extended accretion zone on the NS surface as opposed to the more traditional disc-like boundary layer. Contrary to theory, the SL temperature was seen to increase towards low SL luminosities, while the approximate angular extent had a nearly linear luminosity dependency. Comptonization was not required to adequately fit these spectra. Together with the $\sim70\degr$ upper bound of inclination implied by the lack of eclipses, the best-fitting normalization of the accretion disc blackbody component implies a distance of $\sim10$ kpc, instead of the usually quoted 5 kpc.

\end{abstract}


\section{Introduction}

The Galactic X-ray source GX 9+9 (3A 1728-169, 4U 1728-16) was discovered by a sounding rocket flight performed on 1967 July 7, as reported by \citet{Bradt 68}. An optical counterpart of magnitude \textit{V} = 16.6 with a blue continuum and a large ultraviolet excess was identified by \citet{Davidsen 76}.

GX 9+9 was found to have similar spectral and timing properties to Sco X-1 and other sources that came to be called Low-Mass X-ray Binaries \citep{Mason 76, Parsignault 78}. It was classified as an atoll-type neutron star source in a persistent lower banana state by \citet{Hasinger 89}.

The discovery of an X-ray modulation period from observations performed by the High Energy Astronomy Observatory-1 (\textit{HEAO-1}) in 1977 September was announced \citep{Hertz 86} and refined to 4.19$\pm$0.02 h by \citet{Hertz 88}. The period was interpreted as the binary period of the system. \citet{Schaefer 87, Schaefer 90} found a similar period of 4.198$\pm$0.0094 h in the optical counterpart. This was attributed to the orbital modulation of the X-ray emission reprocessed into visible light in the companion star atmosphere and at the hot spot where the \mbox{accretion} stream hits the accretion disc. \citet{Hertz 88}, as well as \citet{Schaefer 90}, deduced the companion to be an early M-class dwarf, with mass estimates of 0.2--0.45 and $\sim$0.4 M$_{\sun}$, respectively. However, \citet{Reig 03} speculated that the companion may be evolved and earlier than G5, based on aperiodic variability at very low frequencies, similarly to Z sources.

Simultaneous X-ray and optical observations were performed in 1999 August by the Rossi X-ray Timing Explorer (\textit{RXTE}) and the South African Astronomical Observatory (SAAO), respectively \citep{Kong 06}. The orbital period was confirmed in the optical observations, but no corresponding X-ray modulation was found. Neither was there any significant X-ray/optical correlation in the light curves. Several different two-component spectral models consisting of a blackbody and a Comptonized component were successfully fitted to the data. It had already been established previously that the spectrum can not be fitted well with a single-component blackbody, power law or bremsstrahlung model \citep{Schulz 99}. 

\citet{Levine 06} reported a strong increase in the orbital modulation of the X-ray flux in the 2--12 keV band since 2005 January, based on over ten years of \textit{RXTE} All-Sky Monitor (ASM) observations. The peak-to-peak amplitude of the modulation as a fraction of the average intensity was at most 6 per cent from the beginning of the \textit{RXTE} mission in 1996 until 2005 January 19, but was about 18 per cent thereafter to 2005 May 25 and in four intervals between the latter date and 2006 June 9. The modulation was more or less independent of energy. 

The distance of GX 9+9 is not well established, but it is generally considered to be a Galactic bulge object. An estimate of 5 kpc is almost universally used (from \citealt{Christian 97}, based on visual magnitude, X-ray flux during `burst' episodes and an assumed similarity with X1735-44), but occasionally so is 7 kpc \citep{Schaefer 90, Vilhu 07}. 

Recently \citet{Vilhu 07} studied INTErnational Gamma-Ray Astrophysics Laboratory (\textit{INTEGRAL}) and \textit{RXTE} spectra of GX 9+9 from 2003 and 2004 in the framework of the spreading layer theory, a model for the luminous zone formed on a neutron star surface as a continuation of the traditional geometrically thin accretion disc. The spectra were fitted with a model consisting of two blackbody components, one of which is multicolour and originates in the inner accretion disc, while the other is weakly Comptonized and comes from the spreading layer. Estimates of the angular extent of the spreading layer on the stellar surface seemed to agree with the theory, but there was unexpectedly an apparent increase of the spreading layer temperature at low luminosities.

This paper is a continuation of the work of \citet{Vilhu 07}, using a much larger data sample. Section 2 discusses the spreading layer theory, Section 3 presents the observations used in this work and the data reduction methods, Section 4 introduces the spectral model, while Section 5 includes the fitting results, Section 6 some discussion, and Section 7 the final conclusions.

\section{Spreading layers}
\label{SLs}

In non-pulsating neutron star (NS) LMXBs, the magnetic field of the neutron star is weak enough ($\le 10^8$ G) not to affect the accretion flow, and so the accretion disc can extend to the neutron star surface. If the star rotates slower than the Keplerian velocity of the infalling matter, the difference in kinetic energy must be radiated in a boundary layer between the accretion disc and the stellar surface. This energy is of the same order as that radiated in the accretion disc, but comes from a smaller area closer to the star, which should be hotter than the disc and thus produce harder radiation.

\begin{figure}
\begin{center}
\includegraphics[width=0.85\columnwidth]{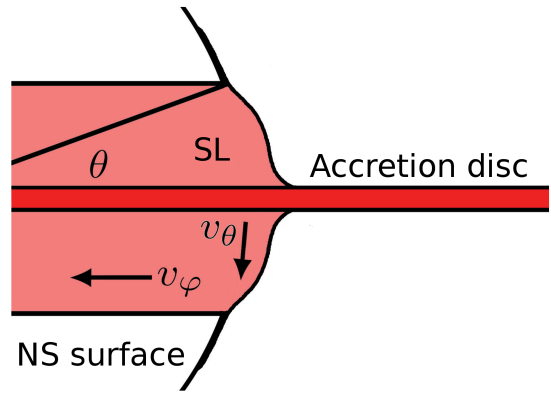}
\caption{\small The spreading layer geometry. Matter flows from the accretion disc onto the neutron star surface and forms the SL, defined as the faster-rotating, radiating part of the spreading flow. $\theta$ is the boundary angle of  this zone, outside which the surface is relatively cool and dark. $v_\theta$ is the latitudinal velocity and $v_\varphi$ the azimuthal (rotational) velocity.}
\label{Fig1}
\end{center}
\end{figure}

The original accretion disc model of \citet{Shakura 73} led to a boundary layer (BL) model (see e.g. \citet{Popham 01} and references therein) that was thin both radially and latitudinally, in which the turbulent friction between the differentially rotating layers of gas was responsible for slowing it down to the stellar rotational velocity. However, \citet{Inogamov 99} introduced a new approach where the principal friction mechanism is the turbulent viscosity between the infalling matter spreading on the stellar surface and the slowly moving dense matter beneath it. After an intermediate bottleneck zone, the latitudinal velocity $v_\theta$ exceeds the radial velocity $v_r$, and the plasma forms a distinct spreading layer (SL), dynamically separate from the accretion disc. Fig. \ref{Fig1} illustrates the basic geometry.
The width of the spreading layer, defined as the area where the matter rotates faster than the underlying surface, increases along with the accretion rate $\dot{M}$, and consequently the luminosity $L$. The best-fitting power-law dependence to these in the numerically integrated model had an index of $\sim0.8$. At a certain rate $\dot{M}_{\mathrm{pole}}$ almost the entire surface is covered by the SL. The radial thickness and its profile also depend on $\dot{M}$; at their maximum in the disc plane they are in the range of 0.2--2 km. Unlike the traditional boundary layer model, where the luminosity has a maximum at the equator, the spreading layer model turns out to have luminosity maxima near the outer edges of the radiating zone, while at the equator there is a local minimum. Radiation from below the spreading flow is reprocessed into the blackbody-like SL spectrum in the optically thick, higher $\dot{M}$ case and Comptonized at very small $\dot{M}$ and $\tau$.

\citet{Suleimanov 06} calculated the spectra produced by a spreading layer at different luminosities, and seen from different inclination angles; the dependence on both of these was slight. The SL spectrum was found to be close to a diluted blackbody. Both the effective temperature $T_{\mathrm{eff}}$ and the hardness factor $f_{\mathrm{c}}$, with which it is multiplied to get the observed colour temperature $T_{\mathrm{c}}$ of the blackbody, increased slightly with SL luminosity. The theory of \citet{Inogamov 99} was developed by taking into account chemical compositions other than pure hydrogen and a general relativity correction to the surface gravity. In the original theory, energy release was assumed to happen mainly in a thin sublayer at the bottom of the SL. Here it was shown that the spectra depend very little on the vertical structure of the SL when the surface density is high (i.e. optically thick case).

A few papers have compared the theory with observations, as well as with the predictions of the traditional boundary layer theory. Evidence supporting the SL scenario has been found, but the superiority of either model is yet to be established conclusively.

\citet{Church 02} compared the SL and BL theories with the results of a survey of LMXBs using the Advanced Satellite for Cosmology and Astrophysics (\textit{ASCA}), which were originally published by \citet{Church 01}. They found good agreement with the former at low luminosities, but at higher luminosities the blackbody emission exceeded predictions by a factor of two to four, suggesting that radial flow dominated the emitting area. 

\citet{Gilfanov 03} and \citet{Revnivtsev 06} used Fourier frequency resolved spectroscopy to study the short term variability of luminous LMXBs. They found that the shapes of the more variable (on $\sim$second--millisecond time scales) hard component were blackbody-like and luminosity-independent, with colour temperatures of $2.4\pm0.1$ keV. This was consistent with the SL model. \citet{Suleimanov 06} also compared their theoretical SL spectra to those of \citet{Revnivtsev 06}, finding in five cases out of six a soft excess below 10 keV. This was interpreted as possibly coming from a classical boundary layer, existing between the accretion disc and the spreading layer.

\section{Observations and data reduction}

The basic data set for this work was formed by all available pointed observations of GX 9+9 by the \textit{INTEGRAL} \citep{Winkler 03} and \textit{RXTE} \citep{Bradt 93} satellites between 2002 and 2007.

For \textit{INTEGRAL}, we used data from both Joint European X-ray Monitors (JEM-X, \citealt{Lund 03}; JEM-X2 before 2004 March, JEM-X1 afterwards) and the \textit{INTEGRAL} Soft Gamma-Ray Imager (ISGRI, \citealt{Lebrun 03}) -- the low energy detector of the Imager on Board the \textit{INTEGRAL} Satellite (IBIS,\citealt{Ubertini 03}). All data were reduced using the Off-line Science Analysis (\textsc{OSA}) version 7 software. To create average spectra, all pointings with more than a few hundred seconds of good JEM-X data and with offsets below $5\degr$ were used. For the individual spectra, to ensure sufficient S/N, we limited the offsets to $4\degr$.

The \textit{RXTE} instruments utilized were Proportional Counter Units (PCUs) 0 and 2 of the Proportional Counter Array (PCA, \citealt{Jahoda 96}), as the rest are off most of the time, using all three Xenon layers, and mostly cluster A, but for a few observations when it was off cluster B of the High Energy X-ray Timing Experiment (HEXTE, \citealt{Rothschild 98}). PCA spectra were extracted from the Standard-2 mode data, using 16 s binning, and HEXTE spectra from the Archive mode data, using 32 s binning to match the dwell times.

Observations over several orbital periods were divided into continuous single-orbit viewings of the source. The time resolution is thus about 3200 seconds.

The Good Time Interval (GTI) criteria were the following: PCUs 0 and 2 must be on, the offset from target less than $0.02\degr$, the elevation above $10\degr$ to eliminate earth occultations and atmospheric influence, the parameter value ELECTRON2 (a measure of background electron contamination for PCU 2, usually identical for the other PCUs) below 0.1 and the time since South Atlantic Anomaly passage more than 10 minutes (or negative). The GTI extensions in the data files themselves were also applied, except for one observation where this resulted in no data at all.

The PCA deadtime correction factors were calculated from the corresponding Standard-1 data. The epoch 5 bright background model was used to create the background data files, from which the PCA background spectra were extracted. The two HEXTE background fields produced by the beamswitching were reduced together.

The final set consisted of 42 JEM-X1+ISGRI, 62 JEM-X2+ISGRI and 92 PCA+HEXTE spectra (196 in total), spanning 2002 May 1 to 2007 July 4.

\section{The model}
\label{model}

The spectral modelling was done with \textsc{Xspec} v.12.4.0, using a similar model as the one in \citet{Vilhu 07}. It essentially consists of two modified blackbody components, one of which represents the accretion disc and the other the spreading layer, plus interstellar absorption. The composite model is defined as ${\tt const*wabs(diskbb + compbb)}$.

LMXB spectral models such as this, based on ${\tt diskbb + compbb}$, have been dubbed \emph{eastern}, in contrast to the \emph{western} models based on a basic blackbody from the boundary layer and a Comptonized continuum from a disc corona, e.g. ${\tt bb + comptt}$ \citep{White 86}; for a comparison see \citet{Paizis 05}.
 
{\tt Compbb} is a Comptonized blackbody model after \citet{Nishimura 86}. This model is used for the spreading layer. There are four model parameters: blackbody temperature k$T$, electron temperature of the Comptonizing hot plasma k$T_{\mathrm{e}}$, optical depth of the plasma $\tau$, and the normalization $N_{\mathrm{SL}}=(R_{\mathrm{km}}/D_{10})^2$, where $R_{\mathrm{km}}$ is the source radius in km (if the source is spherical, which is not the case here; see Section \ref{angle}) and $D_{10}$ is the distance to the source in units of 10 kpc. The Comptonization part is accurate for photon energies up to the (non-relativistic) electron temperature, i.e. $E\la$k$T_{\mathrm{e}}<m_{\mathrm{e}}c^2$, and particularly important when the optical depth is small ($\tau$<3). However, here we set k$T_{\mathrm{e}}$ to equal the blackbody temperature k$T$, as we have found the Comptonization hard to constrain and weak even at best, and assume it to take place on or near the SL surface. \citet{Vilhu 07} found their best-fitting optical thickness to be $0.53$, using k$T_{\mathrm{e}}=10$ keV.

{\tt Diskbb} is a multi-temperature blackbody model for accretion discs, discussed in e.g. \citet{Mitsuda 84} and \citet{Makishima 86}. The model parameters are $T_{\mathrm{in}}$, the temperature at the inner disc radius, and the normalization $N_{\mathrm{Disc}}=(R_{\mathrm{in}}/D_{10})^2\mathrm{cos}\:i$, where $R_{\mathrm{in}}$ is an `apparent' inner disc radius in km,  $D_{10}$ is the distance as in {\tt compbb}, and $i$ is the inclination of the disc normal from the line of sight.

The true inner radius of the disc $R$, which we also assume to be the neutron star and spreading layer radius (ignoring the thickness of the layer), relates to $R_{\mathrm{in}}$ as  $R=R_{\mathrm{in}}\xi\kappa^2$, where $\xi=(3/7)^{1/2}(6/7)^3\approx0.41$ is a factor that arises from a relativistic boundary condition, which causes the disc temperature to reach its maximum outside the inner radius \citep{Kubota 98}, and $\kappa$ is a spectral hardness factor, which can be assumed to stay at about 1.7 \citep{Makishima 00, Shimura 95}. Therefore setting $R_{\mathrm{in}}$ to 10 km roughly corresponds to an actual neutron star radius of 12 km when taking this factor into account.

{\tt Wabs} is an interstellar photoelectric absorption model using Wisconsin cross sections \citep{Morrison 83} and the \citet{Anders 82} relative element abundances. The neutral Hydrogen column density $n_{\mathrm{H}}$ was frozen to $0.196\cdot10^{22}$cm$^{-2}$, calculated from the LAB survey \citep{Kalberla 05} weighted average of seven points within $1\degr$ of GX 9+9, by the HEASARC $n_{\mathrm{H}}$ web tool.

The model was also multiplied by a constant, energy-independent factor ({\tt const}) for instrument cross-calibrations. It was frozen to the deadtime correction factor for the PCA spectra (1.002--1.008), or to unity for the JEM-X spectra, and left free for the other instruments.

\section{Results}

\subsection{Spectral fitting}

The energy bands taken into account were 3--27 keV for PCA, 4--18 keV for JEM-X1 and JEM-X2, 10--27 keV for HEXTE and 12--42 keV for ISGRI; the upper limits were set where the background level typically exceeded the signal for each instrument. 3.0 per cent systematic errors were used for the individual JEM-X+ISGRI spectra, and 2.0 per cent errors for the individual PCA+HEXTE spectra. The fitting and the $1\sigma$ confidence interval calculations were done by the Levenberg-Marquardt method.

\begin{figure}
  \includegraphics[width=\columnwidth]{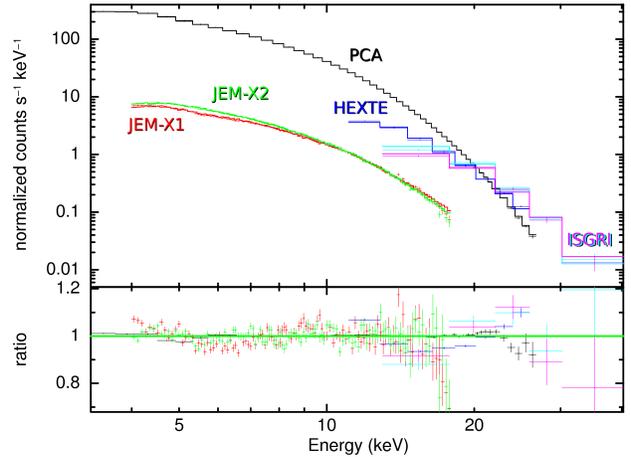}
  \caption{\small Average spectra for the instruments, folded ${\tt const*wabs(diskbb + compbb)}$ model, and data to model ratios.}
\label{Fig2}
\end{figure}

The averaged spectra from all the instruments, fitted simultaneously, are shown in Fig. \ref{Fig2}. In the fitting, the $N_{\mathrm{Disc}}$ parameters for the HEXTE, ISGRI and JEM-X datasets were tied to the PCA $N_{\mathrm{Disc}}$ parameter. This is because we assume the accretion disc to constantly reach the neutron star surface, feeding the optically thick spreading layer and enabling the blackbody-like hard spectral component to exist. This leads to $N_{\mathrm{Disc}}$ being constant, as defined in the previous section. The PCA data were used to establish the value, as they had the tightest statistical error limits.

All other parameters for the HEXTE and ISGRI datasets were tied to their equivalents in the PCA and JEM-X datasets respectively.

\begin{figure}
  \includegraphics[width=\columnwidth]{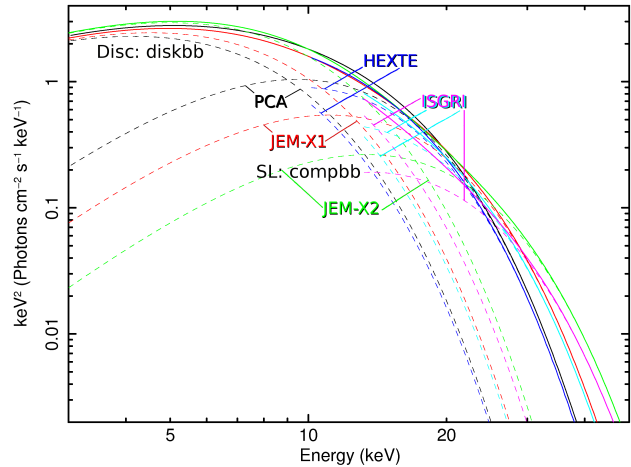}
  \caption{\small Unfolded model corresponding to Fig. \ref{Fig2}. The lower-energy family of component curves represents the accretion disc component (${\tt diskbb}$) and higher-energy family the spreading layer component (${\tt compbb}$).}
\label{Fig3}
\end{figure}

As the JEM-X1+ISGRI and JEM-X2+ISGRI observations are from different times than each other and the PCA+HEXTE observations, the time-variable parameters $T_{\mathrm{in}}$, k$T$ (= k$T_{\mathrm{e}}$), $\tau$, and $N_{\mathrm{SL}}$ for these spectral groups were left independent of each other. 2.5 per cent systematic errors were used for all the average spectra. The best-fitting parameter values are given in Table \ref{averageparams}. As can be seen here, the JEM-X+ISGRI spectra favoured rather different values than the PCA+HEXTE spectra. The unfolded model fitted to the average spectra can be seen in Fig. \ref{Fig3}.

\begin{table}
\begin{center}
\caption{\small Best-fit parameters of the average spectra. $T_{\mathrm{in}}$ and k$T$ are the inner accretion disc and spreading layer temperatures, respectively, while  $N_{\mathrm{SL}}$ is the spreading layer normalization. The accretion disc normalization $N_{\mathrm{Disc}}=36_{-2}^{+1}$ was tied to the PCA parameter for all instruments. The Comptonizing electron temperature k$T_{\mathrm{e}}$ was tied to the SL blackbody temperature. PCA and JEM-X spectra were fitted together with corresponding HEXTE and ISGRI spectra, respectively. The bolometric model (without absorption) flux was $9.7\cdot 10^{-9}$ erg cm$^{-2}$ s$^{-1}$, $2.3\cdot 10^{-9}$ erg cm$^{-2}$ s$^{-1}$ of which was from the {\tt compbb} component. $\chi^2/dof=256.44/277\approx0.93$.}
\label{averageparams}
\begin{tabular}{l|c|c|c|c}
\hline
& $T_{\mathrm{in}}$ [keV] & k$T$, k$T_{\mathrm{e}}$ [keV] & $\tau$ & $N_{\mathrm{SL}}$ \\
\hline
PCA & $1.76_{-0.02}^{+0.01}$ & $2.42_{-0.02}^{+0.02}$ & $0.0_{-0.0}^{+0.1}$ & $5.9_{-0.0}^{+0.9}$ \\
JEM-X1 & $0.97_{-0.02}^{+0.01}$ & $1.691_{-0.011}^{+0.010}$ & $2.18_{-0.11}^{+0.11}$ & $12_{-1}^{+1}$ \\
JEM-X2 & $1.29_{-0.06}^{+0.07}$ & $1.699_{-0.014}^{+0.011}$ & $2.21_{-0.12}^{+0.11}$ & $46_{-4}^{+2}$ \\
\hline
\end{tabular}
\end{center}
\end{table}

\begin{figure*}
\begin{center}
  \includegraphics[width=0.95\textwidth]{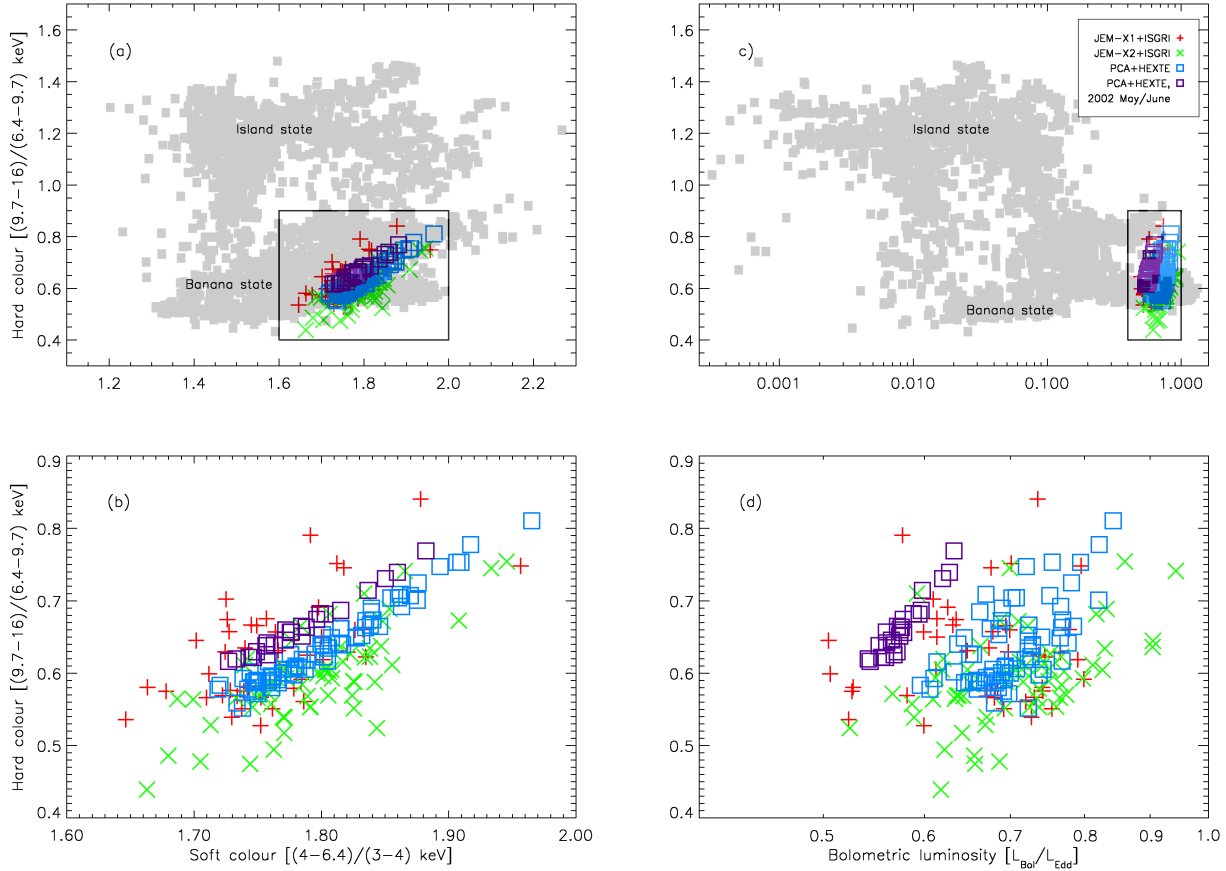}
  \caption{\small (a) A colour-colour diagram of GX 9+9. The grey area is adapted from \citet{Gladstone 07} and represents all the atoll source data; (b) the boxed area of (a); (c) A corresponding colour-luminosity diagram, assuming a distance of 10 kpc and a mass of 1.4 M$_{\sun}$; (d) the boxed area of (c).}
  \label{Fig4}
  \end{center}
\end{figure*}

The individual spectra were then fitted with the same model as for the averaged spectra, but in this case the $N_{\mathrm{Disc}}$ parameter was frozen to the best-fitting value obtained from the averaged spectra. Taking into account the confidence limits ($36_{-2}^{+1}$), the fits were thus done with $N_{\mathrm{Disc}}=34$, $36$ and $37$. The final confidence limits for the other parameters were obtained by summing in quadrature their $1\sigma$ errors in the $N_{\mathrm{Disc}}=36$ fits with the differences in the best-fitting values between the fits using $N_{\mathrm{Disc}}=36$ and those using $N_{\mathrm{Disc}}=34$ and $N_{\mathrm{Disc}}=37$.

The Comptonization optical thickness was hard to constrain for the individual spectra, so as an initial hypothesis it was frozen to zero, as suggested by the average PCA+HEXTE spectra (which were more constraining than the JEM-X+ISGRI spectra). This effectively rendered {\tt compbb} into a basic area-normalized blackbody  ({\tt bbodyrad}). Good fits were obtained for nearly all the individual spectra. For those with $\chi^2/dof>1.2$, 19 out of 195, another fit was made with $\tau$ as a free parameter, to see if it would increase and if the fit would improve. This was not the case for any of the 8 PCA+HEXTE spectra. 5 out of 11 JEM-X+ISGRI spectra gained an increase in $\tau$, but even at best $\chi^2/dof$ only went from 139.17/114 to 132.97/113. It was concluded that partial Comptonization in the SL was not significantly detected.

The results from four occasions of partially overlapping \textit{INTEGRAL} and \textit{RXTE} observations from 2003 October (11 spectra) were seen to be consistent within error.

The means for the free parameters in the individual fits were $T_{\mathrm{in}}=1.787\pm0.005$ keV, k$T=2.50\pm0.02$ keV and $N_{\mathrm{SL}}=5.5\pm0.2$; mean $\chi^2/dof=0.957\pm0.014$. Sample tables of the best-fitting parameters (full tables in the on-line version of this paper) can be found in Appendix \ref{tableappendix}.

\subsection{Fluxes, colours and luminosities}
\label{FCL}

Response-independent, unabsorbed fluxes were integrated over the model spectra over four energy bands: band A = 3--4 keV, band B = 4--6.4 keV, band C = 6.4--9.7 keV and band D = 9.7--16 keV. Hardness ratios were calculated from the fluxes, defining the soft colour as $F_{\mathrm{B}}/F_{\mathrm{A}}$ and the hard colour as $F_{\mathrm{D}}/F_{\mathrm{C}}$, as in \citet{Done 03}, and subsequently \citet{Gladstone 07} and \citet{Vilhu 07}. The colour-colour diagram of Figs. \ref{Fig4}(a) and \ref{Fig4}(b) shows GX 9+9 consistently occupying the banana state, as expected. The typical correlation seen in GX 9+9 is more evident in the PCA+HEXTE results, which had considerably less statistical variance. The earliest PCA+HEXTE observations from 2002 May 1, May 2, June 6 and June 11 form their own somewhat distinct track 
(the purple open squares).    

Bolometric fluxes and {\tt compbb} (SL) component fluxes were calculated, the latter by setting the {\tt diskbb} normalization $N_{\mathrm{Disc}}$ to zero. The {\tt diskbb} fluxes were derived by subtracting the {\tt compbb} fluxes from the corresponding total fluxes.

\begin{figure}
  \includegraphics[width=\columnwidth]{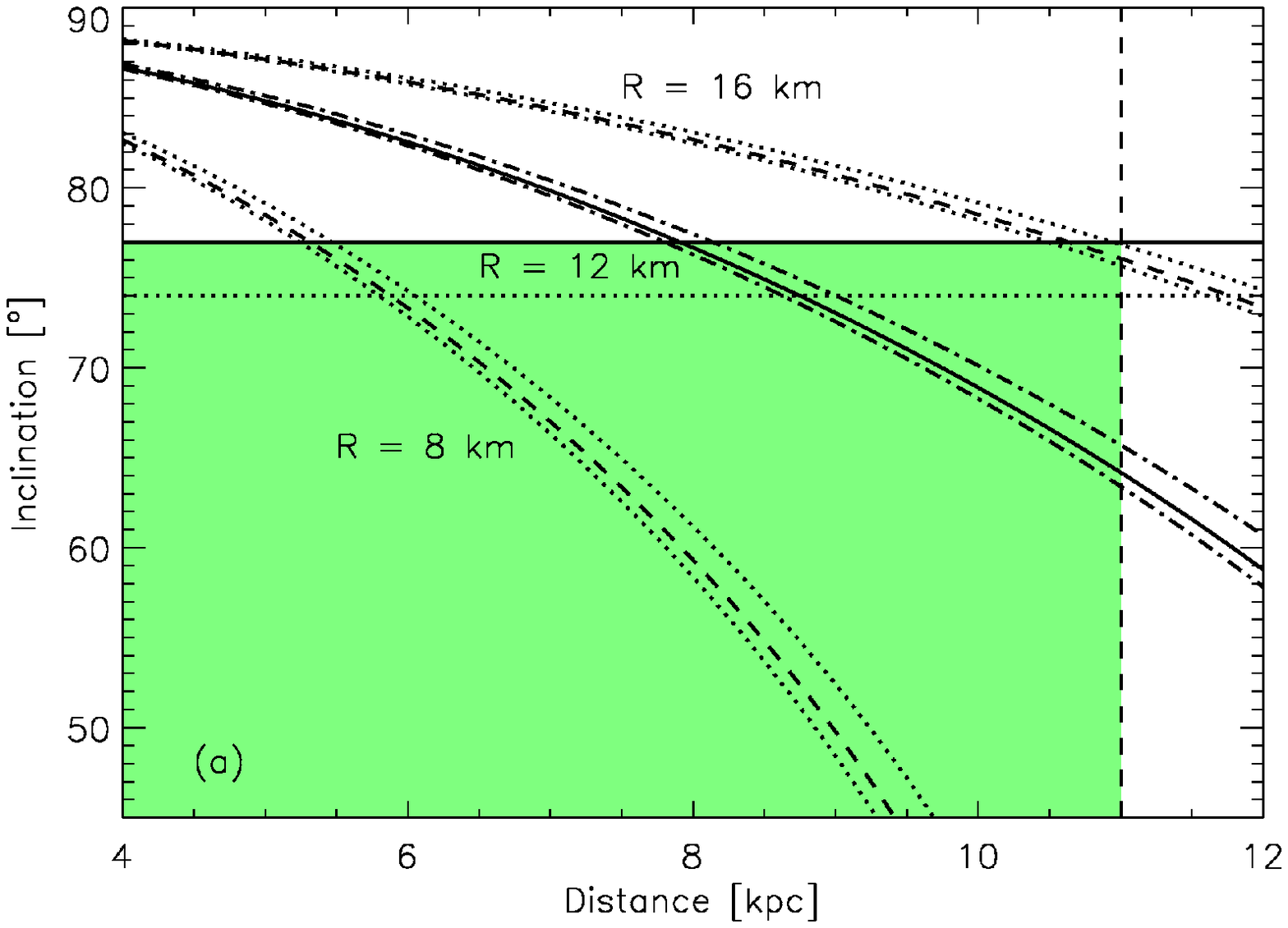}
  \begin{center}
  \includegraphics[width=0.9\columnwidth]{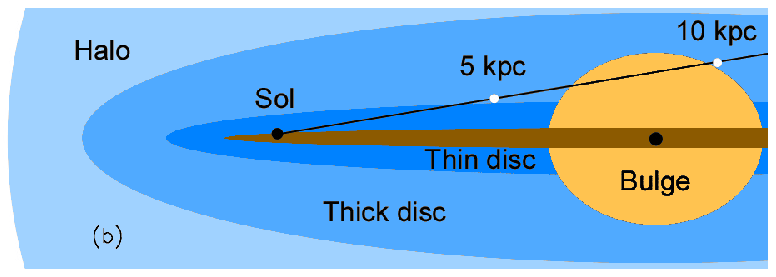}
  \end{center}
  \caption{\small (a) Inclination versus distance for $N_{\mathrm{Disc}}=(R_{\mathrm{in}}/D_{10})^2\mathrm{cos}\:i=36_{-2}^{+1}$ at three different values of $R\approx1.19R_{\mathrm{in}}$ (curves). The horizontal lines are the maximum inclinations in the case of a 0.2 or 0.45 M$_{\sun}$ secondary, $77\degr$ and $74\degr$, while the vertical dashed line denotes the distance of $\sim11$ kpc, where the high end of the luminosity distribution calculated from the model fluxes surpasses the Eddington limit. The region of acceptability is shaded; (b) Suggested locations of GX 9+9 along the line of sight (white dots) and a cross section of the Galactic environment (after \citealt{Buser 00}). GX 9+9 most likely belongs to either the bulge or the thick disc population.}
  \label{Fig5}
\end{figure}

\begin{figure}
  \includegraphics[width=\columnwidth]{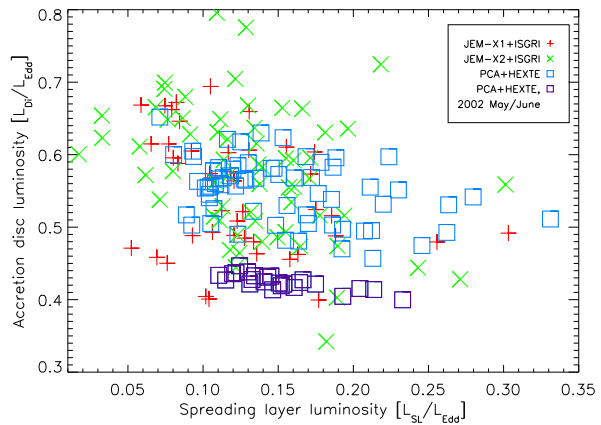}
  \caption{\small Luminosities of the two model components.}
  \label{Fig6}
\end{figure}

To calculate bolometric luminosities from the fluxes, a value has to be assumed for the distance. Let us consider the parameter $N_{\mathrm{Disc}}=(R_{\mathrm{in}}/D_{10})^2\mathrm{cos}\:i=36^{+1}_{-2}$. Using $R_{\mathrm{in}}=10$ km, as discussed previously, and the most generally used value for the distance, 5 kpc, would result in an inclination $i$ of $84.8\degr$. This is unrealistic, as even though the orbital X-ray modulation has increased since early 2005 \citep{Levine 06}, GX 9+9 is not an eclipsing binary. \citet{Paczynski 71} relates the radius of the secondary Roche lobe $r_1$ and the orbital separation $A$ to the masses of the secondary and the neutron star, $M_1$ and $M_2$:
\[
  \frac{r_1}{A}=0.46224\left(\frac{M_1}{M_1+M_2}\right)^{1/3}
\]
The lack of eclipses constrains this to be $<1/(\mathrm{tan}\:i)$, so for a 1.4 M$_{\sun}$ neutron star and a 0.2--0.45 M$_{\sun}$ secondary the inclination upper limit is $74\degr$--$77\degr$. \citet{Schaefer 90} further reasons that the lack of significant dips in the light curve implies an inclination less than $\sim70\degr$. On the other hand, distances larger than 11 kpc are increasingly unlikely, as the binary would be further and further away from the Galactic bulge, and its luminosities increasingly above the Eddington limit $L_{\mathrm{Edd}}$. Therefore for the rest of this paper, we assume a distance of 10 kpc. While being twice the usually quoted value, it has the merit of leaving the inclination just below 70\degr without leading to supercritical luminosities, and is quite reasonable for a Galactic bulge object. Fig. \ref{Fig5} shows the parameter dependences for the best-fitting value and confidence limits, as well as the possible distances along the line of sight in Galactic context. This new distance estimate does put GX 9+9 in a less populated area in the colour-luminosity diagram of Figs. \ref{Fig4}(c) and \ref{Fig4}(d), in the approximate range of 0.5--0.9 $L_{\mathrm{Edd}}$. Fig. \ref{Fig6} shows the relative luminosities of the accretion disc and spreading layer components. There is no clear correlation between these, except for the earliest PCA+HEXTE observations from 2002 May/June, where the accretion disc luminosity was rather constant at a little over 0.4 $L_{\mathrm{Edd}}$. The correlation is smeared out by variations in the total luminosity, in a smaller part due to orbital modulation.

\section{Discussion}

\begin{figure}
  \includegraphics[width=\columnwidth]{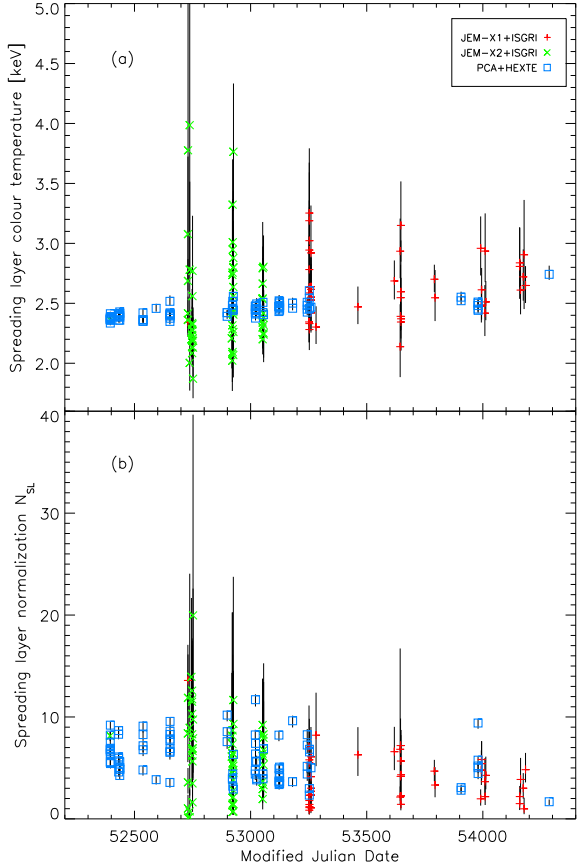}
  \caption{\small Best-fit spreading layer parameter values vs. time (MJD). (a) k$T$; (b) $N_{\mathrm{SL}}$.}
  \label{Fig7}
\end{figure}

Of all the available spreading layer variables, luminosity should be the one most closely correlated with mass accretion rate, the driving force behind observed changes not due to orbital modulation. Therefore the other spreading layer parameters are presented in relation to it. As can be seen from Fig. \ref{Fig7}, k$T$ and $N_{\mathrm{SL}}$ vary considerably within a few hours or days, but are close to constant on longer time scales, with perhaps a slight indication of a rising trend in the former and a decreasing trend in the latter.

\subsection{Spreading layer temperature}

\begin{figure}
  \includegraphics[width=\columnwidth]{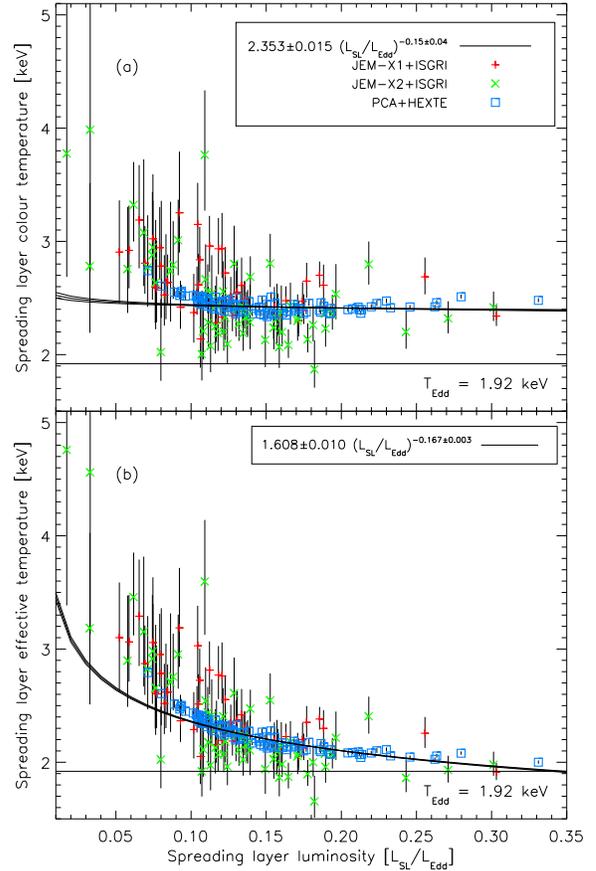}
  \caption{\small (a) Observed spreading layer colour temperature vs. luminosity; (b) effective spreading layer temperature as calculated from (a) by dividing with the hardness factor $f_{\mathrm{c}}$ (Equation \ref{fcformula}). The  \textit{RXTE}/PCA+HEXTE error bars are mostly within the symbols. Also shown are best-fitting power laws ($\chi^2/dof\approx 2.7$) and the Eddington temperature for spherical emission.}
    \label{Fig8}
\end{figure}

In the models of \citet{Suleimanov 06}, the effective SL  temperature $T_{\mathrm{eff}}$ decreases slightly with decreasing SL luminosity. The observed colour temperature (k$T=T_{\mathrm{c}}=f_{\mathrm{c}}T_{\mathrm{eff}}$) results in Fig. \ref{Fig8}(a) show an opposite trend, as they also did in \citet{Vilhu 07}. This may be due to an actual increase of the effective temperature from some low accretion rate factor that has not been properly considered in the theory, or the luminosity dependence of the hardness factor  $f_{\mathrm{c}}=T_{\mathrm{c}}/T_{\mathrm{eff}}$. In the latter case, $f_{\mathrm{c}}$ should have a similar or somewhat stronger low-luminosity increase. The analytical formula for $f_{\mathrm{c}}$ of \citet{Pavlov 91} that was used by \citet{Suleimanov 06} instead gives a strong decrease at luminosities below $0.1\ L_{\mathrm{Edd}}$:
\begin{equation}
  f_{\mathrm{c}}=\left(0.14\ \mathrm{ln}\left(\frac{3+5X}{1-L}\right)+0.59\right)^{-\frac{4}{5}}\left(\frac{3+5X}{1-L}\right)^{\frac{2}{15}}L^{\frac{3}{20}}
  \label{fcformula}
\end{equation}
where $X$ is the hydrogen mass fraction and the luminosity $L$ is expressed in Eddington units. This formula has been successful in describing near-Eddington luminosity \mbox{($>0.9\ L_{\mathrm{Edd}}$)} X-ray burst spectra \citep{Pavlov 91}, but may be incorrect in the low-luminosity domain where it has not been extensively tested. As can be seen in Fig. \ref{Fig8}(b), where the effective temperature as given by Equation (\ref{fcformula}) is shown, this dependency for the hardness factor only serves to highlight the discrepancy between the results and the theory, leading to a clear supercritical temperature increase towards the lower luminosities. The critical Eddington effective temperature $T_{\mathrm{Edd}}$ is determined by the balance between the surface gravity and the radiative acceleration:
\[
  \frac{GM}{R^2\sqrt{1-R_{\mathrm{S}}/R}}=\frac{\sigma_{\mathrm{SB}}T^4_{\mathrm{Edd}}}{c}\sigma_{\mathrm{e}}
\]
\[
  \Rightarrow T_{\mathrm{Edd}}=\sqrt[4]{\frac{GMc}{\sigma_{\mathrm{SB}}\sigma_{\mathrm{e}}R^2\sqrt{1-2GM/c^2R}}}
\]
where $\sigma_{\mathrm{e}}=0.02(1+X)$ m$^2$ kg$^{-1}$ is the electron scattering opacity, X is the hydrogen mass fraction and $\sigma_{\mathrm{SB}}$ is the Stefan-Boltzmann constant \cite{Suleimanov 06}. For solar composition plasma ($X=0.7$), $M=1.4$ M$_{\sun}$ and $R=12000$ m, $T_{\mathrm{Edd}}\approx1.92$ keV. 

\subsection{Spreading layer boundary angle}
\label{angle}

As mentioned in Section \ref{model}, the spreading layer normalization $N_{\mathrm{SL}}$ is proportional to $R_{\mathrm{km}}^2$ and to the inverse of $D_{10}^2$. As the spreading layer is not a spherical source, $R_{\mathrm{km}}\neq R$. To estimate the extent of the SL area to even some degree of accuracy, we need to relate these with a corrective term; this should, in effect, be the ratio between the projection on a plane perpendicular to the line of sight of the visible, luminous SL area $A_{\mathrm{SL}}$ and $\pi R^2$, the corresponding projection of the whole neutron star area.

Here we ignore relativistic light bending, which does actually cause the far hemisphere of a neutron star to be visible up to an angle of $\sim$20--40\degr beyond the classical horizon. Even so, a few further simplifying assumptions have to be made. As in \citet{Vilhu 07}, the spreading layer is approximated by a spherical zone extending from the plane of the accretion disc to a certain boundary angle $\theta$; the accretion disc is opaque and large enough compared to the neutron star to hide the other side of the SL at all inclinations $i<90\degr$. This approximation probably tends to underestimate the upper boundary angle, as according to theory most of the radiation should come from a narrower belt that doesn't reach the equator. But as we can only presume to see the projected area and not the shape of it, or its position on the stellar surface, having two different boundary angles as free parameters might give arbitrary results. The mathematics of the approximation are presented in Appendix \ref{angleappendix}.

The results agree well with those of a simple Monte Carlo simulation. For small values of ($N_{\mathrm{SL}}$, $\theta$), both the approximation and the simulation depend only weakly on the inclination at $i>60\degr$. The approximation becomes less accurate the more of the SL on the other hemisphere is visible.

\begin{figure}
  \includegraphics[width=\columnwidth]{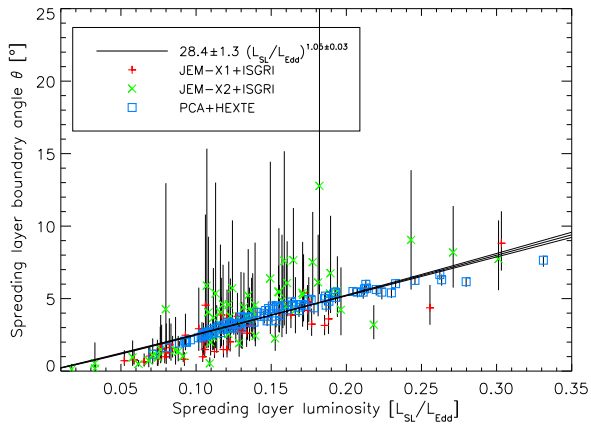}
  \caption{\small Approximate spreading layer boundary angle vs. luminosity and best-fitting power law ($\chi^2/dof\approx1.5$).}
  \label{Fig9}
\end{figure}

The resulting boundary angle values are depicted in Fig. \ref{Fig9} as a function of spreading layer luminosity. The mean of $\theta$ was $(3.67\pm0.14)\degr$. The best-fitting power law dependency is close to linear, as opposed to the theoretically predicted index of $\sim0.8$ \citep{Inogamov 99}. Without proper relativistic corrections, the implications on theory remain unclear.

\section{Conclusions}

We have achieved formally successful fits to a large number of GX 9+9 spectra with a ${\tt diskbb + compbb}$ model, interpreting the ${\tt compbb}$ part as spreading layer emission. The results, however, were somewhat unexpected.

Assuming the model of an accretion disc constantly reaching the neutron star surface is valid, GX 9+9 may be twice as distant as previously thought, of the order of 10 kpc. The distance is suggested by the observed normalization of the soft spectral component identified with the accretion disc ($36_{-2}^{+1}$), together with the upper limit imposed on the system inclination ($\sim70\degr$) by the lack of eclipses. This distance corresponds to a luminosity range of $\sim0.5$--$0.9 L_{\mathrm{Edd}}$. Lack of information on the actual neutron star radius denies us accurate boundaries for the distance.

There was an apparent increase in the temperature of the spreading layer component at low spreading layer luminosities, either due to an effective temperature increase from some low accretion rate factor not considered in the theory, or incorrect theoretical low-luminosity values for the hardness factor $f_{\mathrm{c}}$.

The spreading layer blackbody parameters varied considerably on time scales of a few hours or days, but were close to constant in the long term, the means being k$T\approx2.5$ keV and $N_{\mathrm{SL}}\approx5.5$. The Comptonization parameters were found to be unnecessary, i.e. no hardening to distort the blackbody shape was significantly detected.

Apart from the low-luminosity temperature rise, the results were consistent with spreading layer theory. The relatively stable SL colour temperature of $\sim2.5$ keV is also compatible with the  $2.4\pm0.1$ keV observed in \citet{Gilfanov 03} and \citet{Revnivtsev 06}. The aforementioned short-term studies might have missed a low-luminosity temperature rise, if it did occur in the observed sources. Fourier frequency resolved spectroscopy should also be applied to GX 9+9 at various SL luminosities to further compare the results of the different approaches.

There are also some differences in the results of this and earlier observational studies on the subject. Whereas the observations in \citet{Church 02} had good agreement with the SL theory at low luminosities and trouble at higher ones, in this work the opposite seems to be the case. The observed excess below 10 keV reported by \citet{Suleimanov 06} was also not detected in our study.

\section*{Acknowledgements}

PS acknowledges support from the Academy of Finland, project number 127053, and the Magnus Ehrnrooth foundation, grant number 2008t9. DCH gratefully acknowledges a Fellowship from the Academy of Finland. AP acknowledges the Italian Space Agency financial and programmatic support via contract I/008/07/0.

Based on observations with \emph{INTEGRAL}, an ESA project with instruments and science data centre funded by ESA and member states (especially the PI countries: Denmark, France, Germany, Italy, Switzerland and Spain), the Czech Republic, and Poland and with the participation of Russia and the USA. This research has made use of data obtained from the High Energy Astrophysics Science Archive Research Center (HEASARC), provided by NASA's Goddard Space Flight Center.

\appendix

\section{SL boundary angle approximation}
\label{angleappendix}

\begin{figure*}
\includegraphics[width=0.85\textwidth]{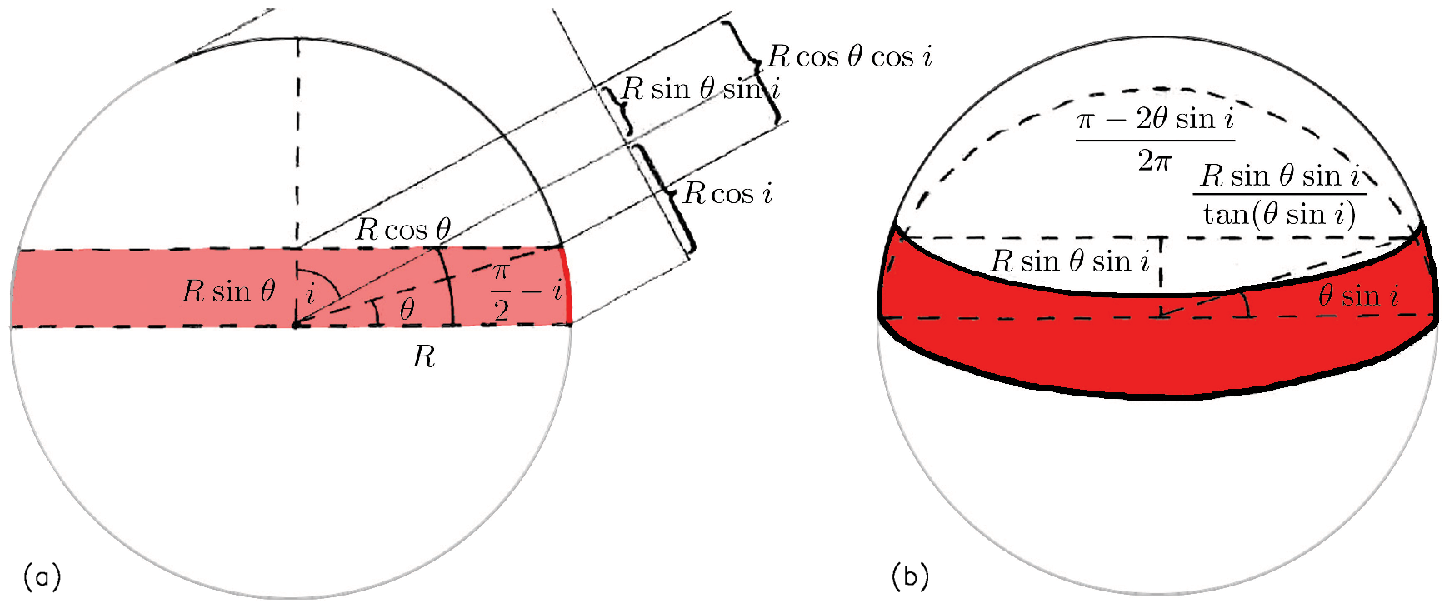}
\caption{\small The geometry of the projected visible spreading layer area approximation. (a) cross-section side view; (b) observer view.}
\label{FigA1}
\end{figure*}

The extreme case closer to the likely inclination of the system (60--70\degr; see Section \ref{FCL}) is the case of $i\approx90\degr$. Viewed nearly edge-on, the accretion disc allows us to see only a semi-circular area of the stellar surface. To get $A_{\mathrm{SL}}$ from this, we have to subtract the polar segment, i.e. the area of the sector whose central angle is $\pi-2\theta_{\mathrm{90}}$, minus the triangle within the spreading layer:
\[
A_{\mathrm{SL}} = \frac{\pi R^2}{2}-\Big(\frac{\pi-2\theta_{\mathrm{90}}}{2\pi}\pi R^2
-2\cdot\frac{1}{2}R^2\:\mathrm{sin}\:\theta_{\mathrm{90}}\:\mathrm{cos}\:\theta_{\mathrm{90}}\Big)
\label{315}
\]
One way to approximate the projected area at intermediate inclinations close to $90\degr$ is to start with the equation above and add a few corrective terms. As depicted in Fig. \ref{FigA1}(a), `vertical' lines and angles are multiplied by the sine of the inclination, while `horizontal' lines are multiplied by the cosine. This doesn't apply to the semimajor axis of the half-ellipse that forms the upper boundary of the SL projection; it has to be derived from the projected angle and the minor cathetus:
\[
\mathrm{tan}(\theta\:\mathrm{sin}\:i)=\frac{R\:\mathrm{sin}\:\theta\:\mathrm{sin}\:i}{x}\Leftrightarrow x=\frac{R\:\mathrm{sin}\:\theta\:\mathrm{sin}\:i}{\mathrm{tan}(\theta\:\mathrm{sin}\:i)}
\]
Adding the area of the lower (`equatorial') half-ellipse and subtracting the area of the upper one, we get an estimate of the shaded area in Fig. \ref{FigA1}(b) that should be viable at reasonably high inclinations and low spreading layer angles:
\[
A_{\mathrm{SL}} = \frac{\pi R^2}{2}-\bigg(\frac{\pi-2\theta\:\mathrm{sin}\:i}{2\pi}\pi R^2
\]
\[
\phantom{A_{\mathrm{SL}} = }
-2\Big(\frac{1}{2}R\:\mathrm{sin}\:\theta\:\mathrm{sin}\:i\:\frac{R\:\mathrm{sin}\:\theta\:\mathrm{sin}\:i}{\mathrm{tan}(\theta\:\mathrm{sin}\:i)}\Big)\bigg)
\]
\[
\phantom{A_{\mathrm{SL}} = }
+\frac{\pi R^2}{2}\:\mathrm{cos}\:i-\frac{\pi R^2}{2}\frac{R\:\mathrm{sin}\:\theta\:\mathrm{sin}\:i}{\mathrm{tan}(\theta\:\mathrm{sin}\:i)}\:\mathrm{cos}\:\theta\:\mathrm{cos}\:i
\]
Now $N_{\mathrm{SL}}$ is solved to single decimal accuracy and compared to observations to find $\theta$:
\[
N_{\mathrm{SL}}=\frac{R^2}{D_{\mathrm{10}}^2}\Big(\frac{\theta\:\mathrm{sin}\:i}{\pi}+\frac{\mathrm{sin}^2\theta\:\mathrm{sin}^2i}{\pi\:\mathrm{tan}(\theta\:\mathrm{sin}\:i)} +\frac{\mathrm{cos}\:i}{2}
\]
\[
\phantom{N_{\mathrm{SL}}=\frac{R^2}{D_{\mathrm{10}}^2}\Big(}
-\frac{\mathrm{sin}\:\theta\:\mathrm{sin}\:i}{2\:\mathrm{tan}(\theta\:\mathrm{sin}\:i)}\mathrm{cos}\:\theta\:\mathrm{cos}\:i\Big)
\label{finalformula2}
\]
$R=12$ (km), $D_{10}=1.0$ (= 10 kpc) and $i=69.0$ (\degr) were used for the calculations (Fig. \ref{Fig9}).
 
\newpage

\section{Best-fitting parameters}
\label{tableappendix}

\newpage

\begin{table*}\small
\begin{center}
\caption{\small Observation ID:s, dates, Modified Julian Dates, best-fitting parameter values, bolometric model fluxes (in $10^{-9}$ erg cm$^{-2}$ s$^{-1}$) and fit statistics of the \textit{RXTE}/PCA+HEXTE spectra.}
\label{RXTEtable}
\begin{tabular}{l|c|c|c|c|c|c|c}
\hline
Obs. ID & MM/DD/YY & MJD & $T_{\mathrm{in}}$ [keV] & k$T$ [keV] & $N_{\mathrm{SL}}$ & $F_{\mathrm{Bol}}$ & $\chi^2/dof$ \\
\hline
 70022-02-01-00 & $ 05/01/02 $ & $ 52395.15 $ & $ 1.66 _{- 0.01 }^{+ 0.03 }$ & $ 2.36 _{- 0.02 }^{+ 0.02 }$ & $ 8.1 _{- 0.5 }^{+ 0.4 }$ & $ 8.73 $ & $ 33.80 / 54 $ \\
 70022-02-01-01G & $ 05/01/02 $ & $ 52395.56 $ & $ 1.68 _{- 0.01 }^{+ 0.03 }$ & $ 2.37 _{- 0.02 }^{+ 0.03 }$ & $ 6.4 _{- 0.6 }^{+ 0.4 }$ & $ 8.42 $ & $ 35.11 / 54 $ \\
 70022-02-01-01G & $ 05/01/02 $ & $ 52395.63 $ & $ 1.68 _{- 0.01 }^{+ 0.03 }$ & $ 2.39 _{- 0.02 }^{+ 0.02 }$ & $ 6.8 _{- 0.5 }^{+ 0.4 }$ & $ 8.67 $ & $ 54.34 / 54 $ \\
 70022-02-01-01G & $ 05/01/02 $ & $ 52395.69 $ & $ 1.69 _{- 0.01 }^{+ 0.03 }$ & $ 2.38 _{- 0.02 }^{+ 0.03 }$ & $ 5.6 _{- 0.5 }^{+ 0.4 }$ & $ 8.36 $ & $ 44.53 / 54 $ \\
 70022-02-01-01G & $ 05/01/02 $ & $ 52395.75 $ & $ 1.66 _{- 0.01 }^{+ 0.03 }$ & $ 2.37 _{- 0.01 }^{+ 0.02 }$ & $ 8.2 _{- 0.5 }^{+ 0.4 }$ & $ 8.81 $ & $ 61.64 / 54 $ \\
 70022-02-01-01G & $ 05/01/02 $ & $ 52395.81 $ & $ 1.68 _{- 0.01 }^{+ 0.03 }$ & $ 2.36 _{- 0.02 }^{+ 0.02 }$ & $ 7.5 _{- 0.5 }^{+ 0.4 }$ & $ 8.79 $ & $ 41.06 / 54 $ \\
 70022-02-01-01 & $ 05/01/02 $ & $ 52395.88 $ & $ 1.68 _{- 0.01 }^{+ 0.03 }$ & $ 2.37 _{- 0.02 }^{+ 0.03 }$ & $ 5.5 _{- 0.5 }^{+ 0.3 }$ & $ 8.15 $ & $ 35.34 / 54 $ \\
 70022-02-01-01 & $ 05/01/02 $ & $ 52395.94 $ & $ 1.68 _{- 0.01 }^{+ 0.03 }$ & $ 2.34 _{- 0.02 }^{+ 0.02 }$ & $ 6.8 _{- 0.5 }^{+ 0.4 }$ & $ 8.46 $ & $ 37.01 / 54 $ \\
 70022-02-01-01 & $ 05/01/02 $ & $ 52396.00 $ & $ 1.68 _{- 0.01 }^{+ 0.03 }$ & $ 2.36 _{- 0.02 }^{+ 0.02 }$ & $ 6.6 _{- 0.5 }^{+ 0.4 }$ & $ 8.48 $ & $ 46.34 / 54 $ \\
 70022-02-01-01 & $ 05/01/02 $ & $ 52396.06 $ & $ 1.68 _{- 0.01 }^{+ 0.03 }$ & $ 2.36 _{- 0.02 }^{+ 0.02 }$ & $ 6.9 _{- 0.5 }^{+ 0.4 }$ & $ 8.51 $ & $ 41.69 / 54 $ \\
 70022-02-01-01 & $ 05/01/02 $ & $ 52396.13 $ & $ 1.66 _{- 0.01 }^{+ 0.03 }$ & $ 2.41 _{- 0.01 }^{+ 0.02 }$ & $ 9.2 _{- 0.5 }^{+ 0.4 }$ & $ 9.33 $ & $ 45.85 / 54 $ \\
 70022-02-01-02 & $ 05/02/02 $ & $ 52396.61 $ & $ 1.67 _{- 0.01 }^{+ 0.03 }$ & $ 2.36 _{- 0.02 }^{+ 0.03 }$ & $ 6.3 _{- 0.5 }^{+ 0.4 }$ & $ 8.26 $ & $ 43.04 / 54 $ \\
 70022-02-01-02 & $ 05/02/02 $ & $ 52396.67 $ & $ 1.69 _{- 0.01 }^{+ 0.03 }$ & $ 2.38 _{- 0.02 }^{+ 0.02 }$ & $ 6.9 _{- 0.5 }^{+ 0.4 }$ & $ 8.76 $ & $ 38.83 / 54 $ \\
 70022-02-01-02 & $ 05/02/02 $ & $ 52396.73 $ & $ 1.69 _{- 0.01 }^{+ 0.03 }$ & $ 2.39 _{- 0.02 }^{+ 0.03 }$ & $ 5.4 _{- 0.5 }^{+ 0.3 }$ & $ 8.26 $ & $ 38.59 / 54 $ \\
 70022-02-05-00 & $ 06/06/02 $ & $ 52431.37 $ & $ 1.68 _{- 0.01 }^{+ 0.03 }$ & $ 2.36 _{- 0.02 }^{+ 0.03 }$ & $ 6.1 _{- 0.5 }^{+ 0.4 }$ & $ 8.33 $ & $ 34.44 / 54 $ \\
 70022-02-05-00 & $ 06/06/02 $ & $ 52431.43 $ & $ 1.68 _{- 0.01 }^{+ 0.03 }$ & $ 2.40 _{- 0.01 }^{+ 0.02 }$ & $ 8.3 _{- 0.5 }^{+ 0.4 }$ & $ 9.14 $ & $ 38.28 / 54 $ \\
 70022-02-05-000 & $ 06/06/02 $ & $ 52431.49 $ & $ 1.67 _{- 0.01 }^{+ 0.03 }$ & $ 2.40 _{- 0.01 }^{+ 0.02 }$ & $ 8.7 _{- 0.5 }^{+ 0.4 }$ & $ 9.25 $ & $ 39.96 / 54 $ \\
 70022-02-05-000 & $ 06/06/02 $ & $ 52431.55 $ & $ 1.69 _{- 0.01 }^{+ 0.03 }$ & $ 2.41 _{- 0.02 }^{+ 0.03 }$ & $ 5.7 _{- 0.5 }^{+ 0.3 }$ & $ 8.46 $ & $ 42.30 / 54 $ \\
 70022-02-05-000 & $ 06/06/02 $ & $ 52431.62 $ & $ 1.69 _{- 0.01 }^{+ 0.03 }$ & $ 2.39 _{- 0.02 }^{+ 0.03 }$ & $ 6.0 _{- 0.5 }^{+ 0.4 }$ & $ 8.52 $ & $ 31.16 / 54 $ \\
 70022-02-05-000 & $ 06/06/02 $ & $ 52431.68 $ & $ 1.69 _{- 0.01 }^{+ 0.03 }$ & $ 2.40 _{- 0.02 }^{+ 0.03 }$ & $ 4.6 _{- 0.5 }^{+ 0.3 }$ & $ 7.99 $ & $ 39.84 / 54 $ \\
 70022-02-06-00 & $ 06/11/02 $ & $ 52436.44 $ & $ 1.70 _{- 0.01 }^{+ 0.03 }$ & $ 2.40 _{- 0.02 }^{+ 0.03 }$ & $ 4.9 _{- 0.5 }^{+ 0.3 }$ & $ 8.19 $ & $ 47.93 / 54 $ \\
 70022-02-06-00 & $ 06/11/02 $ & $ 52436.51 $ & $ 1.69 _{- 0.01 }^{+ 0.03 }$ & $ 2.43 _{- 0.02 }^{+ 0.03 }$ & $ 4.3 _{- 0.4 }^{+ 0.3 }$ & $ 8.02 $ & $ 38.61 / 54 $ \\
 70022-02-06-00 & $ 06/11/02 $ & $ 52436.63 $ & $ 1.70 _{- 0.01 }^{+ 0.03 }$ & $ 2.36 _{- 0.02 }^{+ 0.03 }$ & $ 5.6 _{- 0.5 }^{+ 0.4 }$ & $ 8.37 $ & $ 34.98 / 54 $ \\
 70022-02-06-00 & $ 06/11/02 $ & $ 52436.69 $ & $ 1.70 _{- 0.01 }^{+ 0.03 }$ & $ 2.38 _{- 0.02 }^{+ 0.03 }$ & $ 5.0 _{- 0.5 }^{+ 0.3 }$ & $ 8.25 $ & $ 32.53 / 54 $ \\
 70022-02-06-00 & $ 06/11/02 $ & $ 52436.76 $ & $ 1.71 _{- 0.01 }^{+ 0.03 }$ & $ 2.42 _{- 0.03 }^{+ 0.03 }$ & $ 4.9 _{- 0.5 }^{+ 0.4 }$ & $ 8.42 $ & $ 41.73 / 54 $ \\
 70022-02-01-03 & $ 09/19/02 $ & $ 52536.49 $ & $ 1.75 _{- 0.01 }^{+ 0.03 }$ & $ 2.42 _{- 0.03 }^{+ 0.04 }$ & $ 4.8 _{- 0.6 }^{+ 0.4 }$ & $ 9.04 $ & $ 41.54 / 54 $ \\
 70022-02-02-01 & $ 09/19/02 $ & $ 52536.76 $ & $ 1.71 _{- 0.02 }^{+ 0.03 }$ & $ 2.36 _{- 0.02 }^{+ 0.02 }$ & $ 9.1 _{- 0.6 }^{+ 0.5 }$ & $ 9.88 $ & $ 51.20 / 54 $ \\
 70022-02-02-01 & $ 09/19/02 $ & $ 52536.82 $ & $ 1.74 _{- 0.02 }^{+ 0.03 }$ & $ 2.35 _{- 0.02 }^{+ 0.03 }$ & $ 7.2 _{- 0.6 }^{+ 0.5 }$ & $ 9.50 $ & $ 41.05 / 54 $ \\
 70022-02-02-01 & $ 09/19/02 $ & $ 52536.88 $ & $ 1.73 _{- 0.02 }^{+ 0.03 }$ & $ 2.36 _{- 0.02 }^{+ 0.02 }$ & $ 8.3 _{- 0.6 }^{+ 0.4 }$ & $ 9.77 $ & $ 43.21 / 54 $ \\
 70022-02-02-00 & $ 09/19/02 $ & $ 52536.94 $ & $ 1.74 _{- 0.02 }^{+ 0.03 }$ & $ 2.35 _{- 0.02 }^{+ 0.03 }$ & $ 6.8 _{- 0.6 }^{+ 0.4 }$ & $ 9.39 $ & $ 37.64 / 54 $ \\
 70022-02-03-00 & $ 11/15/02 $ & $ 52593.21 $ & $ 1.76 _{- 0.01 }^{+ 0.03 }$ & $ 2.46 _{- 0.03 }^{+ 0.04 }$ & $ 3.9 _{- 0.5 }^{+ 0.3 }$ & $ 9.00 $ & $ 50.74 / 54 $ \\
 70022-02-04-01 & $ 01/13/03 $ & $ 52652.49 $ & $ 1.78 _{- 0.01 }^{+ 0.03 }$ & $ 2.52 _{- 0.04 }^{+ 0.05 }$ & $ 3.6 _{- 0.5 }^{+ 0.4 }$ & $ 9.33 $ & $ 52.54 / 55 $ \\
 70022-02-04-01 & $ 01/13/03 $ & $ 52652.55 $ & $ 1.76 _{- 0.02 }^{+ 0.03 }$ & $ 2.35 _{- 0.02 }^{+ 0.03 }$ & $ 6.5 _{- 0.7 }^{+ 0.5 }$ & $ 9.60 $ & $ 39.69 / 55 $ \\
 70022-02-04-00 & $ 01/13/03 $ & $ 52652.93 $ & $ 1.75 _{- 0.02 }^{+ 0.03 }$ & $ 2.41 _{- 0.02 }^{+ 0.02 }$ & $ 7.7 _{- 0.6 }^{+ 0.4 }$ & $ 10.17 $ & $ 43.46 / 55 $ \\
 70022-02-04-00 & $ 01/13/03 $ & $ 52653.00 $ & $ 1.77 _{- 0.02 }^{+ 0.03 }$ & $ 2.41 _{- 0.02 }^{+ 0.03 }$ & $ 6.6 _{- 0.6 }^{+ 0.4 }$ & $ 10.11 $ & $ 42.61 / 55 $ \\
 70022-02-04-00 & $ 01/13/03 $ & $ 52653.06 $ & $ 1.75 _{- 0.02 }^{+ 0.03 }$ & $ 2.40 _{- 0.02 }^{+ 0.02 }$ & $ 8.6 _{- 0.6 }^{+ 0.4 }$ & $ 10.43 $ & $ 36.93 / 55 $ \\
 70022-02-04-00 & $ 01/13/03 $ & $ 52653.12 $ & $ 1.73 _{- 0.02 }^{+ 0.03 }$ & $ 2.42 _{- 0.01 }^{+ 0.02 }$ & $ 9.6 _{- 0.6 }^{+ 0.4 }$ & $ 10.62 $ & $ 50.57 / 55 $ \\
 70022-02-04-02 & $ 01/14/03 $ & $ 52653.23 $ & $ 1.76 _{- 0.02 }^{+ 0.03 }$ & $ 2.38 _{- 0.02 }^{+ 0.02 }$ & $ 7.8 _{- 0.6 }^{+ 0.5 }$ & $ 10.16 $ & $ 39.27 / 55 $ \\
 70022-02-04-02 & $ 01/14/03 $ & $ 52653.29 $ & $ 1.75 _{- 0.02 }^{+ 0.03 }$ & $ 2.41 _{- 0.02 }^{+ 0.02 }$ & $ 8.3 _{- 0.6 }^{+ 0.4 }$ & $ 10.35 $ & $ 32.18 / 55 $ \\
 70022-02-04-02 & $ 01/14/03 $ & $ 52653.35 $ & $ 1.76 _{- 0.02 }^{+ 0.03 }$ & $ 2.40 _{- 0.02 }^{+ 0.03 }$ & $ 6.9 _{- 0.6 }^{+ 0.4 }$ & $ 9.95 $ & $ 54.71 / 55 $ \\
 70022-02-07-00 & $ 09/16/03 $ & $ 52898.92 $ & $ 1.79 _{- 0.02 }^{+ 0.03 }$ & $ 2.39 _{- 0.02 }^{+ 0.03 }$ & $ 7.6 _{- 0.7 }^{+ 0.5 }$ & $ 10.67 $ & $ 59.59 / 55 $ \\
 70022-02-07-01 & $ 09/16/03 $ & $ 52898.97 $ & $ 1.75 _{- 0.02 }^{+ 0.03 }$ & $ 2.42 _{- 0.01 }^{+ 0.02 }$ & $ 10.2 _{- 0.6 }^{+ 0.5 }$ & $ 11.13 $ & $ 36.44 / 55 $ \\
 70022-02-07-02 & $ 09/17/03 $ & $ 52899.03 $ & $ 1.78 _{- 0.02 }^{+ 0.03 }$ & $ 2.42 _{- 0.02 }^{+ 0.02 }$ & $ 8.6 _{- 0.6 }^{+ 0.5 }$ & $ 11.08 $ & $ 47.17 / 55 $ \\
 80020-04-01-01 & $ 10/09/03 $ & $ 52921.57 $ & $ 1.82 _{- 0.02 }^{+ 0.03 }$ & $ 2.48 _{- 0.03 }^{+ 0.05 }$ & $ 3.9 _{- 0.6 }^{+ 0.4 }$ & $ 10.21 $ & $ 52.36 / 55 $ \\
 80020-04-01-00 & $ 10/09/03 $ & $ 52921.66 $ & $ 1.82 _{- 0.02 }^{+ 0.03 }$ & $ 2.44 _{- 0.04 }^{+ 0.05 }$ & $ 4.0 _{- 0.7 }^{+ 0.5 }$ & $ 10.10 $ & $ 42.59 / 55 $ \\
 80020-04-02-00 & $ 10/11/03 $ & $ 52923.09 $ & $ 1.82 _{- 0.02 }^{+ 0.03 }$ & $ 2.46 _{- 0.03 }^{+ 0.04 }$ & $ 4.4 _{- 0.6 }^{+ 0.4 }$ & $ 10.32 $ & $ 45.71 / 55 $ \\
 80020-04-02-01 & $ 10/11/03 $ & $ 52923.16 $ & $ 1.81 _{- 0.01 }^{+ 0.03 }$ & $ 2.52 _{- 0.03 }^{+ 0.05 }$ & $ 3.2 _{- 0.5 }^{+ 0.3 }$ & $ 9.72 $ & $ 64.07 / 55 $ \\
 80020-04-03-00 & $ 10/13/03 $ & $ 52925.54 $ & $ 1.83 _{- 0.01 }^{+ 0.03 }$ & $ 2.56 _{- 0.04 }^{+ 0.06 }$ & $ 2.9 _{- 0.5 }^{+ 0.3 }$ & $ 10.19 $ & $ 55.49 / 55 $ \\
 80020-04-04-00 & $ 10/13/03 $ & $ 52925.91 $ & $ 1.83 _{- 0.02 }^{+ 0.03 }$ & $ 2.44 _{- 0.03 }^{+ 0.03 }$ & $ 6.3 _{- 0.7 }^{+ 0.5 }$ & $ 11.29 $ & $ 48.32 / 55 $ \\
 70022-02-08-00 & $ 01/16/04 $ & $ 53020.05 $ & $ 1.81 _{- 0.02 }^{+ 0.03 }$ & $ 2.45 _{- 0.03 }^{+ 0.04 }$ & $ 4.8 _{- 0.6 }^{+ 0.4 }$ & $ 10.23 $ & $ 58.02 / 55 $ \\
 70022-02-08-00 & $ 01/16/04 $ & $ 53020.11 $ & $ 1.80 _{- 0.02 }^{+ 0.03 }$ & $ 2.42 _{- 0.02 }^{+ 0.03 }$ & $ 8.2 _{- 0.7 }^{+ 0.5 }$ & $ 11.30 $ & $ 52.09 / 55 $ \\
 70022-02-08-02 & $ 01/16/04 $ & $ 53020.18 $ & $ 1.82 _{- 0.02 }^{+ 0.03 }$ & $ 2.47 _{- 0.03 }^{+ 0.04 }$ & $ 4.4 _{- 0.6 }^{+ 0.4 }$ & $ 10.44 $ & $ 36.49 / 55 $ \\
 70022-02-07-04 & $ 01/16/04 $ & $ 53020.90 $ & $ 1.76 _{- 0.02 }^{+ 0.03 }$ & $ 2.48 _{- 0.02 }^{+ 0.02 }$ & $ 11.7 _{- 0.7 }^{+ 0.5 }$ & $ 12.42 $ & $ 42.57 / 55 $ \\
 70022-02-07-03 & $ 01/17/04 $ & $ 53021.03 $ & $ 1.81 _{- 0.02 }^{+ 0.03 }$ & $ 2.41 _{- 0.02 }^{+ 0.03 }$ & $ 6.4 _{- 0.7 }^{+ 0.5 }$ & $ 10.74 $ & $ 36.74 / 55 $ \\
 70022-02-07-03 & $ 01/17/04 $ & $ 53021.10 $ & $ 1.79 _{- 0.02 }^{+ 0.03 }$ & $ 2.38 _{- 0.02 }^{+ 0.03 }$ & $ 7.4 _{- 0.7 }^{+ 0.5 }$ & $ 10.66 $ & $ 44.78 / 55 $ \\
 70022-02-07-03 & $ 01/17/04 $ & $ 53021.16 $ & $ 1.81 _{- 0.02 }^{+ 0.03 }$ & $ 2.44 _{- 0.02 }^{+ 0.03 }$ & $ 5.6 _{- 0.6 }^{+ 0.4 }$ & $ 10.66 $ & $ 41.42 / 55 $ \\
 70022-02-07-05 & $ 01/21/04 $ & $ 53025.95 $ & $ 1.82 _{- 0.02 }^{+ 0.03 }$ & $ 2.42 _{- 0.03 }^{+ 0.04 }$ & $ 5.7 _{- 0.7 }^{+ 0.5 }$ & $ 10.71 $ & $ 40.98 / 55 $ \\
 \hline
\end{tabular}
\end{center}
\end{table*}
 
\begin{table*}\small
\begin{center}
\contcaption{\small Observation ID:s, dates, Modified Julian Dates, best-fitting parameter values, bolometric model fluxes (in $10^{-9}$ erg cm$^{-2}$ s$^{-1}$) and fit statistics of the \textit{RXTE}/PCA+HEXTE spectra.}
\begin{tabular}{l|c|c|c|c|c|c|c}
\hline
Obs. ID & MM/DD/YY & MJD & $T_{\mathrm{in}}$ [keV] & k$T$ [keV] & $N_{\mathrm{SL}}$ & $F_{\mathrm{Bol}}$ & $\chi^2/dof$ \\
\hline
 70022-02-08-01 & $ 01/22/04 $ & $ 53026.02 $ & $ 1.80 _{- 0.01 }^{+ 0.03 }$ & $ 2.46 _{- 0.03 }^{+ 0.04 }$ & $ 4.0 _{- 0.5 }^{+ 0.4 }$ & $ 9.85 $ & $ 58.17 / 55 $ \\
 70022-02-08-01 & $ 01/22/04 $ & $ 53026.09 $ & $ 1.81 _{- 0.02 }^{+ 0.03 }$ & $ 2.50 _{- 0.03 }^{+ 0.05 }$ & $ 3.8 _{- 0.6 }^{+ 0.4 }$ & $ 9.99 $ & $ 47.64 / 55 $ \\
 80020-04-05-00 & $ 02/18/04 $ & $ 53053.36 $ & $ 1.85 _{- 0.02 }^{+ 0.03 }$ & $ 2.51 _{- 0.04 }^{+ 0.05 }$ & $ 3.9 _{- 0.6 }^{+ 0.4 }$ & $ 10.87 $ & $ 42.57 / 55 $ \\
 80020-04-05-00 & $ 02/18/04 $ & $ 53053.43 $ & $ 1.81 _{- 0.02 }^{+ 0.03 }$ & $ 2.42 _{- 0.03 }^{+ 0.04 }$ & $ 4.4 _{- 0.6 }^{+ 0.4 }$ & $ 10.05 $ & $ 53.55 / 55 $ \\
 80020-04-06-00 & $ 02/20/04 $ & $ 53055.63 $ & $ 1.81 _{- 0.02 }^{+ 0.03 }$ & $ 2.41 _{- 0.02 }^{+ 0.03 }$ & $ 6.9 _{- 0.7 }^{+ 0.5 }$ & $ 10.92 $ & $ 43.25 / 55 $ \\
 70022-02-09-03 & $ 04/26/04 $ & $ 53121.88 $ & $ 1.79 _{- 0.01 }^{+ 0.03 }$ & $ 2.47 _{- 0.03 }^{+ 0.04 }$ & $ 3.7 _{- 0.5 }^{+ 0.3 }$ & $ 9.51 $ & $ 76.81 / 55 $ \\
 70022-02-09-03 & $ 04/26/04 $ & $ 53121.94 $ & $ 1.81 _{- 0.02 }^{+ 0.03 }$ & $ 2.43 _{- 0.03 }^{+ 0.04 }$ & $ 4.9 _{- 0.6 }^{+ 0.4 }$ & $ 10.22 $ & $ 44.76 / 55 $ \\
 70022-02-09-04 & $ 04/27/04 $ & $ 53122.01 $ & $ 1.80 _{- 0.02 }^{+ 0.03 }$ & $ 2.47 _{- 0.02 }^{+ 0.02 }$ & $ 8.2 _{- 0.6 }^{+ 0.4 }$ & $ 11.52 $ & $ 53.88 / 55 $ \\
 70022-02-09-02 & $ 04/27/04 $ & $ 53122.82 $ & $ 1.80 _{- 0.01 }^{+ 0.03 }$ & $ 2.52 _{- 0.03 }^{+ 0.05 }$ & $ 3.4 _{- 0.5 }^{+ 0.3 }$ & $ 9.66 $ & $ 71.75 / 55 $ \\
 70022-02-09-01 & $ 04/27/04 $ & $ 53122.93 $ & $ 1.80 _{- 0.01 }^{+ 0.03 }$ & $ 2.50 _{- 0.03 }^{+ 0.04 }$ & $ 3.6 _{- 0.5 }^{+ 0.3 }$ & $ 9.76 $ & $ 75.82 / 55 $ \\
 70022-02-09-02 & $ 04/27/04 $ & $ 53122.99 $ & $ 1.81 _{- 0.02 }^{+ 0.03 }$ & $ 2.47 _{- 0.03 }^{+ 0.04 }$ & $ 4.3 _{- 0.6 }^{+ 0.4 }$ & $ 10.16 $ & $ 52.45 / 55 $ \\
 70022-02-09-00 & $ 04/29/04 $ & $ 53124.70 $ & $ 1.80 _{- 0.02 }^{+ 0.03 }$ & $ 2.45 _{- 0.03 }^{+ 0.04 }$ & $ 4.5 _{- 0.6 }^{+ 0.4 }$ & $ 10.00 $ & $ 43.28 / 55 $ \\
 70022-02-09-00 & $ 04/29/04 $ & $ 53124.77 $ & $ 1.81 _{- 0.02 }^{+ 0.03 }$ & $ 2.44 _{- 0.02 }^{+ 0.04 }$ & $ 5.1 _{- 0.6 }^{+ 0.4 }$ & $ 10.36 $ & $ 54.08 / 55 $ \\
 70022-02-09-00 & $ 04/29/04 $ & $ 53124.83 $ & $ 1.79 _{- 0.01 }^{+ 0.03 }$ & $ 2.48 _{- 0.03 }^{+ 0.04 }$ & $ 3.7 _{- 0.5 }^{+ 0.3 }$ & $ 9.58 $ & $ 66.55 / 55 $ \\
 70022-02-09-00 & $ 04/29/04 $ & $ 53124.89 $ & $ 1.80 _{- 0.01 }^{+ 0.03 }$ & $ 2.51 _{- 0.03 }^{+ 0.05 }$ & $ 3.5 _{- 0.5 }^{+ 0.3 }$ & $ 9.71 $ & $ 57.94 / 55 $ \\
 70022-02-09-00 & $ 04/29/04 $ & $ 53124.95 $ & $ 1.80 _{- 0.01 }^{+ 0.03 }$ & $ 2.47 _{- 0.03 }^{+ 0.04 }$ & $ 4.1 _{- 0.5 }^{+ 0.4 }$ & $ 9.88 $ & $ 45.10 / 55 $ \\
 70022-02-10-00 & $ 06/24/04 $ & $ 53180.69 $ & $ 1.82 _{- 0.02 }^{+ 0.03 }$ & $ 2.50 _{- 0.03 }^{+ 0.04 }$ & $ 3.7 _{- 0.5 }^{+ 0.3 }$ & $ 10.09 $ & $ 43.67 / 55 $ \\
 70022-02-09-05 & $ 06/24/04 $ & $ 53180.78 $ & $ 1.78 _{- 0.02 }^{+ 0.03 }$ & $ 2.46 _{- 0.02 }^{+ 0.02 }$ & $ 9.6 _{- 0.7 }^{+ 0.5 }$ & $ 11.71 $ & $ 46.63 / 55 $ \\
 80020-04-07-00 & $ 08/26/04 $ & $ 53243.70 $ & $ 1.85 _{- 0.02 }^{+ 0.03 }$ & $ 2.51 _{- 0.03 }^{+ 0.04 }$ & $ 5.2 _{- 0.7 }^{+ 0.5 }$ & $ 11.45 $ & $ 42.32 / 55 $ \\
 80020-04-07-01 & $ 08/26/04 $ & $ 53243.77 $ & $ 1.83 _{- 0.02 }^{+ 0.03 }$ & $ 2.45 _{- 0.02 }^{+ 0.03 }$ & $ 8.3 _{- 0.8 }^{+ 0.6 }$ & $ 12.10 $ & $ 37.52 / 55 $ \\
 80020-04-07-02 & $ 08/26/04 $ & $ 53243.84 $ & $ 1.83 _{- 0.02 }^{+ 0.03 }$ & $ 2.43 _{- 0.03 }^{+ 0.03 }$ & $ 7.2 _{- 0.8 }^{+ 0.6 }$ & $ 11.56 $ & $ 57.37 / 55 $ \\
 80020-04-07-03 & $ 08/26/04 $ & $ 53243.91 $ & $ 1.85 _{- 0.02 }^{+ 0.03 }$ & $ 2.48 _{- 0.03 }^{+ 0.05 }$ & $ 4.4 _{- 0.7 }^{+ 0.5 }$ & $ 10.95 $ & $ 41.51 / 55 $ \\
 80020-04-08-00 & $ 09/04/04 $ & $ 53252.69 $ & $ 1.83 _{- 0.01 }^{+ 0.03 }$ & $ 2.61 _{- 0.04 }^{+ 0.06 }$ & $ 2.3 _{- 0.4 }^{+ 0.3 }$ & $ 10.03 $ & $ 107.10 / 55 $ \\
 80020-04-08-02 & $ 09/04/04 $ & $ 53252.75 $ & $ 1.84 _{- 0.02 }^{+ 0.03 }$ & $ 2.54 _{- 0.04 }^{+ 0.06 }$ & $ 3.0 _{- 0.5 }^{+ 0.3 }$ & $ 10.29 $ & $ 71.34 / 55 $ \\
 80020-04-08-01 & $ 09/05/04 $ & $ 53253.53 $ & $ 1.83 _{- 0.02 }^{+ 0.03 }$ & $ 2.45 _{- 0.02 }^{+ 0.03 }$ & $ 6.6 _{- 0.7 }^{+ 0.5 }$ & $ 11.32 $ & $ 61.01 / 55 $ \\
 80020-04-08-01 & $ 09/05/04 $ & $ 53253.60 $ & $ 1.82 _{- 0.02 }^{+ 0.03 }$ & $ 2.46 _{- 0.02 }^{+ 0.03 }$ & $ 6.8 _{- 0.7 }^{+ 0.5 }$ & $ 11.34 $ & $ 32.04 / 55 $ \\
 80020-04-09-00 & $ 09/11/04 $ & $ 53259.98 $ & $ 1.86 _{- 0.02 }^{+ 0.03 }$ & $ 2.47 _{- 0.03 }^{+ 0.04 }$ & $ 5.0 _{- 0.7 }^{+ 0.5 }$ & $ 11.33 $ & $ 54.09 / 55 $ \\
 92415-01-01-02 & $ 06/20/06 $ & $ 53906.35 $ & $ 1.77 _{- 0.01 }^{+ 0.03 }$ & $ 2.55 _{- 0.03 }^{+ 0.05 }$ & $ 2.8 _{- 0.4 }^{+ 0.3 }$ & $ 8.93 $ & $ 75.65 / 54 $ \\
 92415-01-01-02 & $ 06/20/06 $ & $ 53906.41 $ & $ 1.76 _{- 0.01 }^{+ 0.03 }$ & $ 2.52 _{- 0.03 }^{+ 0.04 }$ & $ 3.1 _{- 0.4 }^{+ 0.3 }$ & $ 8.78 $ & $ 53.97 / 54 $ \\
 92415-01-02-00 & $ 09/01/06 $ & $ 53979.38 $ & $ 1.82 _{- 0.02 }^{+ 0.03 }$ & $ 2.48 _{- 0.02 }^{+ 0.04 }$ & $ 5.2 _{- 0.6 }^{+ 0.4 }$ & $ 10.74 $ & $ 63.69 / 55 $ \\
 92415-01-02-00 & $ 09/01/06 $ & $ 53979.44 $ & $ 1.78 _{- 0.01 }^{+ 0.03 }$ & $ 2.45 _{- 0.02 }^{+ 0.03 }$ & $ 5.8 _{- 0.5 }^{+ 0.4 }$ & $ 10.11 $ & $ 64.51 / 55 $ \\
 92415-01-02-00 & $ 09/01/06 $ & $ 53979.51 $ & $ 1.79 _{- 0.02 }^{+ 0.03 }$ & $ 2.51 _{- 0.02 }^{+ 0.02 }$ & $ 9.4 _{- 0.6 }^{+ 0.4 }$ & $ 12.11 $ & $ 60.49 / 55 $ \\
 92415-01-02-00 & $ 09/01/06 $ & $ 53979.57 $ & $ 1.83 _{- 0.02 }^{+ 0.03 }$ & $ 2.50 _{- 0.03 }^{+ 0.04 }$ & $ 4.4 _{- 0.5 }^{+ 0.4 }$ & $ 10.60 $ & $ 51.11 / 55 $ \\
 92415-01-02-00 & $ 09/01/06 $ & $ 53979.63 $ & $ 1.77 _{- 0.01 }^{+ 0.03 }$ & $ 2.44 _{- 0.02 }^{+ 0.03 }$ & $ 5.0 _{- 0.5 }^{+ 0.4 }$ & $ 9.66 $ & $ 56.65 / 55 $ \\
 93406-09-01-00 & $ 07/04/07 $ & $ 54285.73 $ & $ 1.87 _{- 0.01 }^{+ 0.03 }$ & $ 2.74 _{- 0.05 }^{+ 0.07 }$ & $ 1.7 _{- 0.3 }^{+ 0.2 }$ & $ 10.66 $ & $ 89.47 / 55 $ \\
 \hline
\end{tabular}
\end{center}
\end{table*}

\begin{table*}\small
\begin{center}
\caption{\small Science Window ID:s, dates, Modified Julian Dates, best-fitting parameter values, bolometric model fluxes (in $10^{-9}$ erg cm$^{-2}$ s$^{-1}$) and fit statistics of the \textit{INTEGRAL}/JEM-X2+ISGRI spectra.}
\label{J2table}
\begin{tabular}{l|c|c|c|c|c|c|c}
\hline
ScW ID & MM/DD/YY & MJD & $T_{\mathrm{in}}$ [keV] & k$T$ [keV] & $N_{\mathrm{SL}}$ & $F_{\mathrm{Bol}}$ & $\chi^2/dof$ \\
\hline
5600760010 & $ 03/31/03 $ & $ 52729.80 $ & $ 1.75 _{- 0.04 }^{+ 0.05 }$ & $ 2.34 _{- 0.09 }^{+ 0.10 }$ & $ 13.6 _{- 3.0 }^{+ 3.5 }$ & $ 11.72 $ & $ 126.25 / 114 $ \\
5600770010 & $ 03/31/03 $ & $ 52729.82 $ & $ 1.77 _{- 0.04 }^{+ 0.05 }$ & $ 2.36 _{- 0.14 }^{+ 0.17 }$ & $ 8.4 _{- 2.8 }^{+ 3.8 }$ & $ 10.47 $ & $ 109.34 / 114 $ \\
5601000010 & $ 04/01/03 $ & $ 52730.33 $ & $ 1.89 _{- 0.03 }^{+ 0.04 }$ & $ 3.08 _{- 0.47 }^{+ 0.65 }$ & $ 1.0 _{- 0.7 }^{+ 1.3 }$ & $ 10.82 $ & $ 101.08 / 114 $ \\
5601010010 & $ 04/01/03 $ & $ 52730.35 $ & $ 1.82 _{- 0.02 }^{+ 0.04 }$ & $ 2.69 _{- 0.16 }^{+ 0.18 }$ & $ 3.6 _{- 1.1 }^{+ 1.4 }$ & $ 10.65 $ & $ 101.08 / 114 $ \\
5601020010 & $ 04/01/03 $ & $ 52730.38 $ & $ 1.81 _{- 0.05 }^{+ 0.05 }$ & $ 2.42 _{- 0.12 }^{+ 0.14 }$ & $ 11.9 _{- 3.4 }^{+ 4.3 }$ & $ 12.68 $ & $ 114.03 / 114 $ \\
5601090010 & $ 04/01/03 $ & $ 52730.54 $ & $ 1.84 _{- 0.03 }^{+ 0.02 }$ & $ 3.78 _{- 1.09 }^{+ 3.86 }$ & $ 0.1 _{- 0.1 }^{+ 0.5 }$ & $ 9.11 $ & $ 124.69 / 114 $ \\
5900290010 & $ 04/08/03 $ & $ 52737.81 $ & $ 1.86 _{- 0.02 }^{+ 0.02 }$ & $ 3.99 _{- 1.25 }^{+ 1.28 }$ & $ 0.2 _{- 0.2 }^{+ 0.5 }$ & $ 9.67 $ & $ 128.63 / 114 $ \\
5900300010 & $ 04/08/03 $ & $ 52737.84 $ & $ 1.88 _{- 0.03 }^{+ 0.03 }$ & $ 2.78 _{- 0.59 }^{+ 0.73 }$ & $ 0.7 _{- 0.6 }^{+ 2.2 }$ & $ 10.12 $ & $ 108.84 / 114 $ \\
5900310010 & $ 04/08/03 $ & $ 52737.86 $ & $ 1.77 _{- 0.10 }^{+ 0.06 }$ & $ 2.00 _{- 0.23 }^{+ 0.43 }$ & $ 8.9 _{- 6.7 }^{+ 15.1 }$ & $ 9.17 $ & $ 115.37 / 114 $ \\
6100920010 & $ 04/16/03 $ & $ 52745.26 $ & $ 1.70 _{- 0.07 }^{+ 0.06 }$ & $ 2.20 _{- 0.15 }^{+ 0.18 }$ & $ 13.9 _{- 5.4 }^{+ 7.7 }$ & $ 10.13 $ & $ 124.72 / 114 $ \\
6200730010 & $ 04/18/03 $ & $ 52748.00 $ & $ 1.74 _{- 0.07 }^{+ 0.06 }$ & $ 2.29 _{- 0.24 }^{+ 0.33 }$ & $ 6.9 _{- 3.8 }^{+ 6.8 }$ & $ 9.13 $ & $ 126.12 / 114 $ \\
6200740010 & $ 04/19/03 $ & $ 52748.02 $ & $ 1.77 _{- 0.05 }^{+ 0.05 }$ & $ 2.13 _{- 0.12 }^{+ 0.14 }$ & $ 11.5 _{- 4.1 }^{+ 5.5 }$ & $ 10.23 $ & $ 119.04 / 114 $ \\
6200750010 & $ 04/19/03 $ & $ 52748.04 $ & $ 1.69 _{- 0.06 }^{+ 0.06 }$ & $ 2.32 _{- 0.13 }^{+ 0.15 }$ & $ 12.6 _{- 3.8 }^{+ 5.1 }$ & $ 10.31 $ & $ 120.18 / 114 $ \\
\hline
\end{tabular}
\end{center}
\end{table*}

\begin{table*}\small
\begin{center}
\contcaption{\small Science Window ID:s, dates, Modified Julian Dates, best-fitting parameter values, bolometric model fluxes (in $10^{-9}$ erg cm$^{-2}$ s$^{-1}$) and fit statistics of the \textit{INTEGRAL}/JEM-X2+ISGRI spectra.}
\begin{tabular}{l|c|c|c|c|c|c|c}
\hline
ScW ID & MM/DD/YY & MJD & $T_{\mathrm{in}}$ [keV] & k$T$ [keV] & $N_{\mathrm{SL}}$ & $F_{\mathrm{Bol}}$ & $\chi^2/dof$ \\
\hline
6200890010 & $ 04/19/03 $ & $ 52748.35 $ & $ 1.73 _{- 0.05 }^{+ 0.05 }$ & $ 2.08 _{- 0.11 }^{+ 0.14 }$ & $ 11.7 _{- 4.2 }^{+ 5.8 }$ & $ 9.41 $ & $ 98.48 / 114 $ \\
6200900010 & $ 04/19/03 $ & $ 52748.37 $ & $ 1.75 _{- 0.03 }^{+ 0.04 }$ & $ 2.24 _{- 0.10 }^{+ 0.11 }$ & $ 8.3 _{- 2.3 }^{+ 2.9 }$ & $ 9.56 $ & $ 117.49 / 114 $ \\
6200910010 & $ 04/19/03 $ & $ 52748.41 $ & $ 1.81 _{- 0.04 }^{+ 0.05 }$ & $ 2.34 _{- 0.18 }^{+ 0.24 }$ & $ 6.2 _{- 2.9 }^{+ 4.1 }$ & $ 10.28 $ & $ 113.41 / 114 $ \\
6200920010 & $ 04/19/03 $ & $ 52748.43 $ & $ 1.73 _{- 0.06 }^{+ 0.06 }$ & $ 2.23 _{- 0.16 }^{+ 0.20 }$ & $ 10.3 _{- 4.3 }^{+ 6.3 }$ & $ 9.78 $ & $ 125.70 / 114 $ \\
6300330010 & $ 04/20/03 $ & $ 52749.83 $ & $ 1.75 _{- 0.04 }^{+ 0.04 }$ & $ 2.33 _{- 0.20 }^{+ 0.25 }$ & $ 5.6 _{- 2.6 }^{+ 4.1 }$ & $ 9.02 $ & $ 108.19 / 114 $ \\
6300340010 & $ 04/20/03 $ & $ 52749.85 $ & $ 1.73 _{- 0.04 }^{+ 0.04 }$ & $ 2.19 _{- 0.16 }^{+ 0.19 }$ & $ 6.9 _{- 2.9 }^{+ 4.3 }$ & $ 8.65 $ & $ 115.41 / 114 $ \\
6300350010 & $ 04/20/03 $ & $ 52749.87 $ & $ 1.76 _{- 0.03 }^{+ 0.04 }$ & $ 2.56 _{- 0.19 }^{+ 0.24 }$ & $ 3.5 _{- 1.4 }^{+ 1.9 }$ & $ 9.15 $ & $ 115.12 / 114 $ \\
6300360010 & $ 04/20/03 $ & $ 52749.90 $ & $ 1.79 _{- 0.03 }^{+ 0.04 }$ & $ 2.77 _{- 0.35 }^{+ 0.46 }$ & $ 1.6 _{- 0.9 }^{+ 1.7 }$ & $ 8.98 $ & $ 113.74 / 114 $ \\
6300500010 & $ 04/21/03 $ & $ 52750.21 $ & $ 1.71 _{- 0.05 }^{+ 0.04 }$ & $ 2.20 _{- 0.17 }^{+ 0.21 }$ & $ 6.9 _{- 2.9 }^{+ 4.6 }$ & $ 8.35 $ & $ 128.63 / 114 $ \\
6300510010 & $ 04/21/03 $ & $ 52750.23 $ & $ 1.77 _{- 0.04 }^{+ 0.04 }$ & $ 2.27 _{- 0.15 }^{+ 0.19 }$ & $ 6.7 _{- 2.6 }^{+ 3.7 }$ & $ 9.62 $ & $ 106.33 / 114 $ \\
6400090010 & $ 04/23/03 $ & $ 52752.13 $ & $ 1.60 _{- 0.14 }^{+ 0.10 }$ & $ 1.87 _{- 0.16 }^{+ 0.26 }$ & $ 20.0 _{- 11.9 }^{+ 19.7 }$ & $ 7.73 $ & $ 139.17 / 114 $ \\
6400100010 & $ 04/23/03 $ & $ 52752.15 $ & $ 1.76 _{- 0.05 }^{+ 0.05 }$ & $ 2.19 _{- 0.15 }^{+ 0.20 }$ & $ 7.9 _{- 3.4 }^{+ 5.0 }$ & $ 9.48 $ & $ 103.03 / 114 $ \\
6400110010 & $ 04/23/03 $ & $ 52752.17 $ & $ 1.74 _{- 0.10 }^{+ 0.07 }$ & $ 2.13 _{- 0.24 }^{+ 0.34 }$ & $ 9.8 _{- 5.9 }^{+ 12.8 }$ & $ 9.37 $ & $ 112.21 / 114 $ \\
11901120010 & $ 10/06/03 $ & $ 52918.99 $ & $ 1.81 _{- 0.07 }^{+ 0.05 }$ & $ 2.21 _{- 0.25 }^{+ 0.35 }$ & $ 6.1 _{- 3.8 }^{+ 8.2 }$ & $ 9.89 $ & $ 103.87 / 114 $ \\
12000820010 & $ 10/09/03 $ & $ 52921.28 $ & $ 1.82 _{- 0.08 }^{+ 0.06 }$ & $ 2.02 _{- 0.25 }^{+ 0.47 }$ & $ 6.4 _{- 5.1 }^{+ 13.8 }$ & $ 9.69 $ & $ 126.41 / 114 $ \\
12000830010 & $ 10/09/03 $ & $ 52921.30 $ & $ 1.78 _{- 0.09 }^{+ 0.07 }$ & $ 2.08 _{- 0.23 }^{+ 0.42 }$ & $ 8.2 _{- 5.9 }^{+ 12.1 }$ & $ 9.46 $ & $ 115.50 / 114 $ \\
12000980010 & $ 10/09/03 $ & $ 52921.64 $ & $ 1.87 _{- 0.04 }^{+ 0.04 }$ & $ 2.74 _{- 0.45 }^{+ 0.61 }$ & $ 2.1 _{- 1.5 }^{+ 3.6 }$ & $ 10.86 $ & $ 130.45 / 114 $ \\
12000990010 & $ 10/09/03 $ & $ 52921.66 $ & $ 1.82 _{- 0.04 }^{+ 0.04 }$ & $ 2.43 _{- 0.21 }^{+ 0.26 }$ & $ 4.5 _{- 2.0 }^{+ 3.2 }$ & $ 10.15 $ & $ 104.14 / 114 $ \\
12001000010 & $ 10/09/03 $ & $ 52921.68 $ & $ 1.80 _{- 0.07 }^{+ 0.06 }$ & $ 2.32 _{- 0.23 }^{+ 0.31 }$ & $ 7.3 _{- 3.9 }^{+ 7.2 }$ & $ 10.50 $ & $ 94.34 / 114 $ \\
12100060010 & $ 10/10/03 $ & $ 52922.53 $ & $ 1.72 _{- 0.06 }^{+ 0.05 }$ & $ 2.09 _{- 0.18 }^{+ 0.23 }$ & $ 8.7 _{- 4.4 }^{+ 7.4 }$ & $ 8.68 $ & $ 113.37 / 114 $ \\
12100070010 & $ 10/10/03 $ & $ 52922.55 $ & $ 1.82 _{- 0.02 }^{+ 0.03 }$ & $ 3.32 _{- 0.32 }^{+ 0.38 }$ & $ 0.7 _{- 0.3 }^{+ 0.4 }$ & $ 9.34 $ & $ 118.75 / 114 $ \\
12100080010 & $ 10/10/03 $ & $ 52922.58 $ & $ 1.85 _{- 0.03 }^{+ 0.03 }$ & $ 2.76 _{- 0.39 }^{+ 0.55 }$ & $ 1.3 _{- 0.9 }^{+ 1.8 }$ & $ 9.86 $ & $ 108.45 / 114 $ \\
12100310010 & $ 10/11/03 $ & $ 52923.09 $ & $ 1.85 _{- 0.04 }^{+ 0.04 }$ & $ 2.41 _{- 0.18 }^{+ 0.23 }$ & $ 5.4 _{- 2.3 }^{+ 3.4 }$ & $ 11.07 $ & $ 95.06 / 114 $ \\
12100320010 & $ 10/11/03 $ & $ 52923.11 $ & $ 1.87 _{- 0.05 }^{+ 0.05 }$ & $ 2.28 _{- 0.22 }^{+ 0.32 }$ & $ 5.5 _{- 3.3 }^{+ 5.4 }$ & $ 11.21 $ & $ 134.08 / 114 $ \\
12100330010 & $ 10/11/03 $ & $ 52923.13 $ & $ 1.91 _{- 0.02 }^{+ 0.03 }$ & $ 2.95 _{- 0.35 }^{+ 0.41 }$ & $ 1.3 _{- 0.7 }^{+ 1.2 }$ & $ 11.41 $ & $ 106.52 / 114 $ \\
12100410010 & $ 10/11/03 $ & $ 52923.31 $ & $ 1.86 _{- 0.02 }^{+ 0.03 }$ & $ 3.01 _{- 0.31 }^{+ 0.36 }$ & $ 1.5 _{- 0.6 }^{+ 1.1 }$ & $ 10.61 $ & $ 105.37 / 114 $ \\
12200060010 & $ 10/13/03 $ & $ 52925.53 $ & $ 1.87 _{- 0.04 }^{+ 0.04 }$ & $ 2.64 _{- 0.41 }^{+ 0.55 }$ & $ 2.1 _{- 1.4 }^{+ 3.7 }$ & $ 10.69 $ & $ 150.95 / 114 $ \\
12200070010 & $ 10/13/03 $ & $ 52925.54 $ & $ 1.90 _{- 0.03 }^{+ 0.04 }$ & $ 2.88 _{- 0.39 }^{+ 0.56 }$ & $ 1.5 _{- 1.0 }^{+ 1.7 }$ & $ 11.28 $ & $ 100.77 / 114 $ \\
12200080010 & $ 10/13/03 $ & $ 52925.56 $ & $ 1.79 _{- 0.09 }^{+ 0.07 }$ & $ 2.07 _{- 0.19 }^{+ 0.29 }$ & $ 11.7 _{- 7.0 }^{+ 12.1 }$ & $ 10.22 $ & $ 104.56 / 114 $ \\
12200210010 & $ 10/13/03 $ & $ 52925.85 $ & $ 1.89 _{- 0.05 }^{+ 0.05 }$ & $ 2.41 _{- 0.25 }^{+ 0.33 }$ & $ 5.2 _{- 2.8 }^{+ 5.2 }$ & $ 11.76 $ & $ 99.51 / 114 $ \\
12200220010 & $ 10/13/03 $ & $ 52925.87 $ & $ 1.96 _{- 0.03 }^{+ 0.04 }$ & $ 2.80 _{- 0.26 }^{+ 0.34 }$ & $ 2.8 _{- 1.4 }^{+ 2.0 }$ & $ 13.33 $ & $ 122.82 / 114 $ \\
12200230010 & $ 10/13/03 $ & $ 52925.89 $ & $ 1.91 _{- 0.03 }^{+ 0.04 }$ & $ 2.56 _{- 0.21 }^{+ 0.27 }$ & $ 3.8 _{- 1.8 }^{+ 2.5 }$ & $ 12.18 $ & $ 113.65 / 114 $ \\
12200240010 & $ 10/13/03 $ & $ 52925.92 $ & $ 1.88 _{- 0.04 }^{+ 0.05 }$ & $ 2.43 _{- 0.16 }^{+ 0.20 }$ & $ 6.4 _{- 2.5 }^{+ 3.5 }$ & $ 12.23 $ & $ 144.03 / 114 $ \\
12200380010 & $ 10/14/03 $ & $ 52926.23 $ & $ 1.97 _{- 0.02 }^{+ 0.03 }$ & $ 3.76 _{- 0.49 }^{+ 0.57 }$ & $ 0.7 _{- 0.3 }^{+ 0.6 }$ & $ 13.34 $ & $ 126.28 / 114 $ \\
12200390010 & $ 10/14/03 $ & $ 52926.25 $ & $ 1.86 _{- 0.05 }^{+ 0.05 }$ & $ 2.26 _{- 0.14 }^{+ 0.18 }$ & $ 9.3 _{- 3.7 }^{+ 5.2 }$ & $ 11.97 $ & $ 100.62 / 114 $ \\
12200400010 & $ 10/14/03 $ & $ 52926.27 $ & $ 1.93 _{- 0.03 }^{+ 0.04 }$ & $ 2.80 _{- 0.18 }^{+ 0.20 }$ & $ 4.8 _{- 1.5 }^{+ 2.1 }$ & $ 13.90 $ & $ 134.50 / 114 $ \\
16400250010 & $ 02/16/04 $ & $ 53051.65 $ & $ 1.86 _{- 0.05 }^{+ 0.05 }$ & $ 2.53 _{- 0.22 }^{+ 0.27 }$ & $ 6.4 _{- 2.7 }^{+ 4.6 }$ & $ 12.27 $ & $ 99.48 / 114 $ \\
16400260010 & $ 02/16/04 $ & $ 53051.67 $ & $ 1.81 _{- 0.05 }^{+ 0.05 }$ & $ 2.30 _{- 0.18 }^{+ 0.23 }$ & $ 8.2 _{- 3.7 }^{+ 5.5 }$ & $ 10.88 $ & $ 142.47 / 114 $ \\
16400400010 & $ 02/16/04 $ & $ 53051.98 $ & $ 1.83 _{- 0.04 }^{+ 0.04 }$ & $ 2.54 _{- 0.26 }^{+ 0.34 }$ & $ 3.7 _{- 1.8 }^{+ 3.2 }$ & $ 10.42 $ & $ 113.52 / 114 $ \\
16400410010 & $ 02/17/04 $ & $ 53052.01 $ & $ 1.83 _{- 0.04 }^{+ 0.04 }$ & $ 2.34 _{- 0.17 }^{+ 0.21 }$ & $ 6.1 _{- 2.6 }^{+ 3.7 }$ & $ 10.67 $ & $ 110.75 / 114 $ \\
16400420010 & $ 02/17/04 $ & $ 53052.03 $ & $ 1.86 _{- 0.03 }^{+ 0.04 }$ & $ 2.66 _{- 0.23 }^{+ 0.31 }$ & $ 2.9 _{- 1.4 }^{+ 2.0 }$ & $ 10.92 $ & $ 122.15 / 114 $ \\
16400430010 & $ 02/17/04 $ & $ 53052.05 $ & $ 1.80 _{- 0.04 }^{+ 0.05 }$ & $ 2.20 _{- 0.13 }^{+ 0.15 }$ & $ 9.2 _{- 3.4 }^{+ 4.6 }$ & $ 10.56 $ & $ 134.41 / 114 $ \\
16400440010 & $ 02/17/04 $ & $ 53052.07 $ & $ 1.90 _{- 0.03 }^{+ 0.04 }$ & $ 2.79 _{- 0.30 }^{+ 0.39 }$ & $ 1.9 _{- 1.0 }^{+ 1.7 }$ & $ 11.32 $ & $ 102.48 / 114 $ \\
16500640010 & $ 02/20/04 $ & $ 53055.60 $ & $ 1.87 _{- 0.03 }^{+ 0.04 }$ & $ 2.42 _{- 0.16 }^{+ 0.20 }$ & $ 5.3 _{- 2.1 }^{+ 2.8 }$ & $ 11.37 $ & $ 90.45 / 114 $ \\
16500660010 & $ 02/20/04 $ & $ 53055.66 $ & $ 1.88 _{- 0.03 }^{+ 0.04 }$ & $ 2.81 _{- 0.21 }^{+ 0.26 }$ & $ 3.3 _{- 1.3 }^{+ 1.7 }$ & $ 12.05 $ & $ 125.40 / 114 $ \\
16500670010 & $ 02/20/04 $ & $ 53055.68 $ & $ 1.83 _{- 0.04 }^{+ 0.04 }$ & $ 2.32 _{- 0.13 }^{+ 0.16 }$ & $ 8.0 _{- 2.8 }^{+ 3.6 }$ & $ 11.31 $ & $ 111.32 / 114 $ \\
16500680010 & $ 02/20/04 $ & $ 53055.70 $ & $ 1.83 _{- 0.05 }^{+ 0.05 }$ & $ 2.38 _{- 0.19 }^{+ 0.23 }$ & $ 6.7 _{- 2.9 }^{+ 4.7 }$ & $ 11.09 $ & $ 121.00 / 114 $ \\
16500820010 & $ 02/21/04 $ & $ 53056.02 $ & $ 1.83 _{- 0.06 }^{+ 0.05 }$ & $ 2.24 _{- 0.19 }^{+ 0.26 }$ & $ 8.3 _{- 4.3 }^{+ 7.0 }$ & $ 11.06 $ & $ 85.04 / 114 $ \\
16500830010 & $ 02/21/04 $ & $ 53056.04 $ & $ 1.85 _{- 0.06 }^{+ 0.05 }$ & $ 2.25 _{- 0.24 }^{+ 0.36 }$ & $ 6.1 _{- 3.8 }^{+ 7.3 }$ & $ 10.87 $ & $ 93.07 / 114 $ \\
\hline
\end{tabular}
\end{center}
\end{table*}

\begin{table*}\small
\begin{center}
\caption{\small Science Window ID:s, dates, Modified Julian Dates, best-fitting parameter values, bolometric model fluxes (in $10^{-9}$ erg cm$^{-2}$ s$^{-1}$) and fit statistics of the \textit{INTEGRAL}/JEM-X1+ISGRI spectra.}
\label{J1table}
\begin{tabular}{l|c|c|c|c|c|c|c}
\hline
ScW ID & MM/DD/YY & MJD & $T_{\mathrm{in}}$ [keV] & k$T$ [keV] & $N_{\mathrm{SL}}$ & $F_{\mathrm{Bol}}$ & $\chi^2/dof$ \\
\hline
23100020010 & $ 09/03/04 $ & $ 53251.51 $ & $ 1.72 _{- 0.05 }^{+ 0.05 }$ & $ 2.34 _{- 0.17 }^{+ 0.19 }$ & $ 7.0 _{- 2.6 }^{+ 4.1 }$ & $ 9.04 $ & $ 104.80 / 114 $ \\
23100240010 & $ 09/04/04 $ & $ 53252.41 $ & $ 1.83 _{- 0.04 }^{+ 0.04 }$ & $ 2.78 _{- 0.46 }^{+ 0.58 }$ & $ 1.8 _{- 1.1 }^{+ 3.1 }$ & $ 9.96 $ & $ 109.38 / 114 $ \\
23100250010 & $ 09/04/04 $ & $ 53252.42 $ & $ 1.83 _{- 0.03 }^{+ 0.04 }$ & $ 2.63 _{- 0.32 }^{+ 0.43 }$ & $ 2.3 _{- 1.3 }^{+ 2.6 }$ & $ 9.90 $ & $ 106.35 / 114 $ \\
23100260010 & $ 09/04/04 $ & $ 53252.44 $ & $ 1.85 _{- 0.02 }^{+ 0.03 }$ & $ 3.19 _{- 0.43 }^{+ 0.48 }$ & $ 0.9 _{- 0.4 }^{+ 0.9 }$ & $ 10.03 $ & $ 113.03 / 114 $ \\
23100390010 & $ 09/04/04 $ & $ 53252.73 $ & $ 1.84 _{- 0.04 }^{+ 0.04 }$ & $ 2.28 _{- 0.17 }^{+ 0.22 }$ & $ 5.8 _{- 2.6 }^{+ 4.0 }$ & $ 10.61 $ & $ 129.92 / 114 $ \\
23100400010 & $ 09/04/04 $ & $ 53252.75 $ & $ 1.81 _{- 0.03 }^{+ 0.04 }$ & $ 2.33 _{- 0.15 }^{+ 0.20 }$ & $ 5.5 _{- 2.3 }^{+ 3.1 }$ & $ 10.09 $ & $ 121.47 / 114 $ \\
23100410010 & $ 09/04/04 $ & $ 53252.77 $ & $ 1.85 _{- 0.03 }^{+ 0.04 }$ & $ 2.59 _{- 0.30 }^{+ 0.42 }$ & $ 2.3 _{- 1.4 }^{+ 2.5 }$ & $ 10.20 $ & $ 128.01 / 114 $ \\
23100420010 & $ 09/04/04 $ & $ 53252.80 $ & $ 1.84 _{- 0.02 }^{+ 0.02 }$ & $ 3.25 _{- 0.35 }^{+ 0.54 }$ & $ 1.1 _{- 0.6 }^{+ 0.8 }$ & $ 10.28 $ & $ 105.60 / 114 $ \\
23100430010 & $ 09/04/04 $ & $ 53252.82 $ & $ 1.88 _{- 0.02 }^{+ 0.03 }$ & $ 2.94 _{- 0.28 }^{+ 0.36 }$ & $ 1.4 _{- 0.7 }^{+ 1.0 }$ & $ 10.93 $ & $ 139.58 / 114 $ \\
23100440010 & $ 09/04/04 $ & $ 53252.86 $ & $ 1.89 _{- 0.03 }^{+ 0.03 }$ & $ 3.02 _{- 0.41 }^{+ 0.57 }$ & $ 1.2 _{- 0.7 }^{+ 1.3 }$ & $ 10.94 $ & $ 109.34 / 114 $ \\
23300990010 & $ 09/11/04 $ & $ 53259.84 $ & $ 1.89 _{- 0.02 }^{+ 0.03 }$ & $ 2.92 _{- 0.33 }^{+ 0.40 }$ & $ 1.1 _{- 0.5 }^{+ 0.9 }$ & $ 10.72 $ & $ 139.64 / 114 $ \\
23301000010 & $ 09/11/04 $ & $ 53259.86 $ & $ 1.91 _{- 0.03 }^{+ 0.04 }$ & $ 2.62 _{- 0.21 }^{+ 0.27 }$ & $ 3.0 _{- 1.4 }^{+ 1.9 }$ & $ 11.78 $ & $ 135.98 / 114 $ \\
23301010010 & $ 09/11/04 $ & $ 53259.88 $ & $ 1.85 _{- 0.03 }^{+ 0.04 }$ & $ 2.42 _{- 0.15 }^{+ 0.18 }$ & $ 6.1 _{- 2.2 }^{+ 2.9 }$ & $ 11.29 $ & $ 131.46 / 114 $ \\
23301020010 & $ 09/11/04 $ & $ 53259.90 $ & $ 1.88 _{- 0.03 }^{+ 0.04 }$ & $ 2.55 _{- 0.21 }^{+ 0.28 }$ & $ 4.1 _{- 1.9 }^{+ 2.8 }$ & $ 11.65 $ & $ 141.51 / 114 $ \\
23301030010 & $ 09/11/04 $ & $ 53259.93 $ & $ 1.87 _{- 0.02 }^{+ 0.03 }$ & $ 2.64 _{- 0.19 }^{+ 0.24 }$ & $ 2.3 _{- 1.0 }^{+ 1.3 }$ & $ 10.77 $ & $ 128.15 / 114 $ \\
23301040010 & $ 09/11/04 $ & $ 53259.96 $ & $ 1.89 _{- 0.03 }^{+ 0.04 }$ & $ 2.53 _{- 0.27 }^{+ 0.35 }$ & $ 2.7 _{- 1.5 }^{+ 2.7 }$ & $ 11.12 $ & $ 147.51 / 114 $ \\
24100160010 & $ 10/03/04 $ & $ 53281.83 $ & $ 1.82 _{- 0.04 }^{+ 0.05 }$ & $ 2.30 _{- 0.14 }^{+ 0.17 }$ & $ 8.2 _{- 3.0 }^{+ 4.2 }$ & $ 10.98 $ & $ 147.18 / 114 $ \\
30100390010 & $ 04/01/05 $ & $ 53461.98 $ & $ 1.84 _{- 0.03 }^{+ 0.04 }$ & $ 2.47 _{- 0.14 }^{+ 0.17 }$ & $ 6.3 _{- 2.1 }^{+ 2.7 }$ & $ 11.47 $ & $ 94.78 / 114 $ \\
35300700010 & $ 09/05/05 $ & $ 53618.98 $ & $ 1.74 _{- 0.05 }^{+ 0.05 }$ & $ 2.69 _{- 0.15 }^{+ 0.17 }$ & $ 6.6 _{- 1.8 }^{+ 2.4 }$ & $ 10.84 $ & $ 99.69 / 114 $ \\
36200270010 & $ 10/01/05 $ & $ 53644.01 $ & $ 1.82 _{- 0.02 }^{+ 0.03 }$ & $ 2.93 _{- 0.23 }^{+ 0.27 }$ & $ 2.1 _{- 0.8 }^{+ 1.0 }$ & $ 10.24 $ & $ 102.56 / 114 $ \\
36200280010 & $ 10/01/05 $ & $ 53644.04 $ & $ 1.75 _{- 0.08 }^{+ 0.06 }$ & $ 2.14 _{- 0.25 }^{+ 0.35 }$ & $ 6.8 _{- 4.3 }^{+ 9.9 }$ & $ 8.83 $ & $ 78.05 / 114 $ \\
36200780010 & $ 10/02/05 $ & $ 53645.62 $ & $ 1.78 _{- 0.03 }^{+ 0.04 }$ & $ 2.39 _{- 0.12 }^{+ 0.14 }$ & $ 7.2 _{- 2.1 }^{+ 2.7 }$ & $ 10.31 $ & $ 99.60 / 114 $ \\
36200790010 & $ 10/02/05 $ & $ 53645.65 $ & $ 1.84 _{- 0.04 }^{+ 0.04 }$ & $ 2.55 _{- 0.23 }^{+ 0.28 }$ & $ 4.2 _{- 1.8 }^{+ 3.0 }$ & $ 10.86 $ & $ 125.73 / 114 $ \\
36300250010 & $ 10/03/05 $ & $ 53646.99 $ & $ 1.71 _{- 0.03 }^{+ 0.04 }$ & $ 2.60 _{- 0.32 }^{+ 0.43 }$ & $ 2.3 _{- 1.2 }^{+ 2.3 }$ & $ 7.76 $ & $ 146.54 / 114 $ \\
36300260010 & $ 10/04/05 $ & $ 53647.02 $ & $ 1.74 _{- 0.03 }^{+ 0.04 }$ & $ 2.34 _{- 0.14 }^{+ 0.16 }$ & $ 5.7 _{- 1.9 }^{+ 2.5 }$ & $ 9.03 $ & $ 98.72 / 114 $ \\
36300270010 & $ 10/04/05 $ & $ 53647.06 $ & $ 1.82 _{- 0.02 }^{+ 0.03 }$ & $ 3.15 _{- 0.30 }^{+ 0.37 }$ & $ 1.4 _{- 0.6 }^{+ 0.9 }$ & $ 10.00 $ & $ 121.13 / 114 $ \\
36300520010 & $ 10/04/05 $ & $ 53647.84 $ & $ 1.67 _{- 0.05 }^{+ 0.05 }$ & $ 2.37 _{- 0.26 }^{+ 0.34 }$ & $ 4.3 _{- 2.3 }^{+ 4.4 }$ & $ 7.46 $ & $ 107.92 / 114 $ \\
41100500010 & $ 02/25/06 $ & $ 53791.18 $ & $ 1.77 _{- 0.02 }^{+ 0.03 }$ & $ 2.70 _{- 0.11 }^{+ 0.12 }$ & $ 4.7 _{- 1.0 }^{+ 1.1 }$ & $ 10.34 $ & $ 118.26 / 114 $ \\
41200490010 & $ 02/28/06 $ & $ 53794.25 $ & $ 1.66 _{- 0.03 }^{+ 0.03 }$ & $ 2.55 _{- 0.19 }^{+ 0.23 }$ & $ 3.3 _{- 1.2 }^{+ 1.7 }$ & $ 7.44 $ & $ 100.42 / 114 $ \\
47800210010 & $ 09/12/06 $ & $ 53990.66 $ & $ 1.78 _{- 0.02 }^{+ 0.03 }$ & $ 2.96 _{- 0.23 }^{+ 0.27 }$ & $ 2.0 _{- 0.7 }^{+ 0.9 }$ & $ 9.36 $ & $ 100.76 / 114 $ \\
47900190010 & $ 09/15/06 $ & $ 53993.57 $ & $ 1.72 _{- 0.03 }^{+ 0.04 }$ & $ 2.48 _{- 0.14 }^{+ 0.17 }$ & $ 4.8 _{- 1.5 }^{+ 1.9 }$ & $ 8.83 $ & $ 107.07 / 114 $ \\
47900540010 & $ 09/16/06 $ & $ 53994.70 $ & $ 1.72 _{- 0.03 }^{+ 0.04 }$ & $ 2.48 _{- 0.11 }^{+ 0.13 }$ & $ 5.8 _{- 1.4 }^{+ 1.8 }$ & $ 9.22 $ & $ 124.78 / 114 $ \\
47900550010 & $ 09/16/06 $ & $ 53994.73 $ & $ 1.75 _{- 0.04 }^{+ 0.04 }$ & $ 2.61 _{- 0.16 }^{+ 0.18 }$ & $ 5.4 _{- 1.6 }^{+ 2.2 }$ & $ 9.97 $ & $ 105.55 / 114 $ \\
48400280010 & $ 09/30/06 $ & $ 54008.82 $ & $ 1.75 _{- 0.03 }^{+ 0.04 }$ & $ 2.42 _{- 0.19 }^{+ 0.23 }$ & $ 3.7 _{- 1.5 }^{+ 2.3 }$ & $ 8.57 $ & $ 100.68 / 114 $ \\
48400640010 & $ 10/01/06 $ & $ 54009.98 $ & $ 1.75 _{- 0.03 }^{+ 0.03 }$ & $ 2.94 _{- 0.26 }^{+ 0.31 }$ & $ 2.2 _{- 0.8 }^{+ 1.2 }$ & $ 8.98 $ & $ 113.51 / 114 $ \\
48500590010 & $ 10/04/06 $ & $ 54012.71 $ & $ 1.77 _{- 0.03 }^{+ 0.04 }$ & $ 2.51 _{- 0.15 }^{+ 0.17 }$ & $ 4.3 _{- 1.4 }^{+ 1.8 }$ & $ 9.55 $ & $ 118.25 / 114 $ \\
53400340010 & $ 02/27/07 $ & $ 54158.59 $ & $ 1.72 _{- 0.02 }^{+ 0.03 }$ & $ 2.81 _{- 0.27 }^{+ 0.33 }$ & $ 1.5 _{- 0.6 }^{+ 1.0 }$ & $ 7.78 $ & $ 112.08 / 114 $ \\
53400430010 & $ 02/27/07 $ & $ 54158.88 $ & $ 1.81 _{- 0.02 }^{+ 0.02 }$ & $ 2.84 _{- 0.20 }^{+ 0.29 }$ & $ 2.2 _{- 0.9 }^{+ 1.0 }$ & $ 9.93 $ & $ 99.68 / 114 $ \\
53500630010 & $ 03/03/07 $ & $ 54162.49 $ & $ 1.74 _{- 0.03 }^{+ 0.04 }$ & $ 2.61 _{- 0.20 }^{+ 0.25 }$ & $ 3.9 _{- 1.5 }^{+ 2.1 }$ & $ 9.04 $ & $ 118.71 / 114 $ \\
53900470010 & $ 03/14/07 $ & $ 54173.99 $ & $ 1.76 _{- 0.03 }^{+ 0.03 }$ & $ 2.72 _{- 0.20 }^{+ 0.23 }$ & $ 3.0 _{- 1.0 }^{+ 1.5 }$ & $ 9.31 $ & $ 113.42 / 114 $ \\
54000690010 & $ 03/18/07 $ & $ 54177.61 $ & $ 1.73 _{- 0.03 }^{+ 0.03 }$ & $ 2.91 _{- 0.45 }^{+ 0.46 }$ & $ 1.0 _{- 0.5 }^{+ 1.1 }$ & $ 7.72 $ & $ 139.95 / 114 $ \\
54200760010 & $ 03/24/07 $ & $ 54183.87 $ & $ 1.66 _{- 0.03 }^{+ 0.04 }$ & $ 2.65 _{- 0.15 }^{+ 0.16 }$ & $ 4.8 _{- 1.3 }^{+ 1.6 }$ & $ 8.50 $ & $ 130.58 / 114 $ \\
\hline
\end{tabular}
\end{center}
\end{table*}

\bsp
\label{lastpage}

\end{document}